\documentclass[preprint,aps,showpacs]{revtex4}
\usepackage{graphicx,epsf}
\usepackage{lscape}
\usepackage{citesort}
\begin{document}
\title{{\Large{\bf Rare Semileptonic $B_{s}$ Decays  to $\eta$ and $\eta'$
mesons in QCD }}}

\author{\small K. Azizi$^1$ \footnote{e-mail: kazizi @ dogus.edu.tr},
\small R. Khosravi$^2$ \footnote {e-mail: khosravi.reza @ gmail.com},
\small F. Falahati$^2$ \footnote {e-mail: falahati@shirazu.ac.ir}}

\affiliation{$^1$Physics Division, Faculty of Arts and Sciences, Do\u
gu\c s  University, Ac\i badem-Kad\i k$\ddot{o}$y, 34722 Istanbul,
Turkey\\
$^2$Physics Department, Shiraz University, Shiraz 71454,
Iran}

\begin{abstract}
We analyze the rare semileptonic $B_s \to (\eta, \eta') l^+ l^-$,
$(l=e, \mu, \tau)$ and $B_s \to (\eta, \eta') \nu \bar{\nu}$
transitions probing  the $\bar s s$ content of the $\eta$ and
$\eta'$  mesons  via three--point QCD sum rules.  We calculate
responsible form factors for these transitions in full theory.
Using the obtained form factors, we also estimate the related
branching fractions and longitudinal lepton polarization
asymmetries. Our results are in a good consistency with the
predictions of the other existing nonperturbative approaches.
\end{abstract}


\maketitle

\section{Introduction}
Among  $B$ mesons, the $B_s$ has been received special attention,
since experimentally it is expected that an abundant number of
$B_s$ will be produced at LHCb. This will provide  possibility to
study  properties of the this meson and its various decay
channels. The first evidence for $B_s$ production at the
$\Upsilon(5S)$ peak  was found by the CLEO
collaboration~\cite{CLEO05,CLEO06}.  Recently, the Belle
Collaboration  measured the branching ratios of the $B_s\to
J/\psi\phi$ transition as well as the  $B_s\to J/\psi\eta$ decay
via the $\eta\to \gamma\gamma$ and $\eta \to \pi^+\pi^0\pi^-$
channels  to reconstruct the $\eta $ meson \cite{Belle}.

Semileptonic decays of the $B_s$ to the $\eta$ and $\eta'$,
induced by the rare flavor changing neutral current (FCNC)
transition of $b\to sl^+l^-$ and  $b\to s\nu \bar{\nu}$ are
crucial framework to restrict the  SM parameters. They can provide
possibility to extract  the elements of the Cabbibo-
Kobayashi-Maskawa (CKM) matrix and search for origin of the CP and
T violations. As these transitions occur at the lowest order
through one-loop penguin diagrams, they are good context to search
for new physics effects beyond the SM. Looking for supersymmetric
particles \cite{susy}, light dark matter \cite{dmat} and  fourth
generation of  quarks is possible via these transitions. These
transitions are also useful to study structures of   the $\eta$
and $\eta'$  mesons.

In the present work, we analyze the semileptonic $B_s \to (\eta,
\eta') l^+ l^-/\nu \bar{\nu}$ decays considering also the $\bar s
s$ content of the $\eta$ and $\eta'$  mesons in the framework of
the three point QCD sum rules. Here, we consider also the mixing
between the $\eta$ and $\eta'$ states with  a single mixing angle
\cite{FKS,DP} as:
\begin{eqnarray}
|\eta\rangle &=&  \cos \, \varphi |\eta_q\rangle- {\sin} \, \varphi  |\eta_s\rangle \nonumber \\
|\eta'\rangle &=& { \sin} \, \varphi  |\eta_q\rangle+{\cos} \,
\varphi |\eta_s\rangle  \,\,\,\, . \label{mixing}
\end{eqnarray}
where, in the quark favor (QF) basis (for more details  see for instance \cite{HMC,CCD}),
\begin{eqnarray}
|\eta_q \rangle &=&{1 \over \sqrt{2} } \left( |\bar
{u} u\rangle +|\bar{d} d\rangle\right), \nonumber \\
|\eta_s \rangle &=& |\bar{s} s\rangle \,\, . \label{etaqs}
\end{eqnarray}
The decay constants of $\bar q q$ and $\bar s s$ parts are defined
in terms of the pion decay constant as \cite{FKS}:
\begin{equation}
f_q  =  (1.02\pm 0.02)f_\pi,\qquad f_s = (1.34\pm 0.06)f_\pi.
\end{equation}
We will use the mixing angle  $\varphi=\big( 41.5 \pm 0.3_{stat} \pm 0.7_{syst}
\pm0.6_{th} \big )^\circ$  \cite{KLOE},  which   has recently been obtained by
the KLOE Collaboration in QF basis via measuring the ratio $\displaystyle{
\Gamma(\phi \to \eta^\prime  \gamma) \over \Gamma(\phi \to \eta
\gamma)}$.
In the QF basis with the single mixing angle, the form factors of $B_s\to\eta(\eta')$
transitions are defined in terms of the form factors $B_s\to\eta_s$ as:
\begin{eqnarray}\label{Fee}
f_i^{B_s\to\eta(\eta')} = -\sin\varphi \left(\cos\varphi \right)
f_i^{B_s\to\eta_s},
\end{eqnarray}
 and their branching fractions are also related
to the branching ratio of $B_s\to\eta_s$ as follows:
\begin{eqnarray}\label{Gee}
{\rm BR}\left\{ B_s\to\eta(\eta')l^+l^-\right\} = \sin^2\varphi \left(\cos^2\varphi \right)
{\rm BR}\left\{B_s\to\eta_s l^+ l^-\right\}.
\end{eqnarray}

The paper is organized as follows: sum rules for  form factors
responsible for considered transitions are obtained in Section II.
Section III is devoted to the  numerical analysis of the form
factors, branching ratios and  longitudinal lepton polarization asymmetries
 as well as our discussions. In this
section, we also compare the obtained results with the existing
predictions of the other non-perturbative approaches.

\section{ QCD sum rules for transition form factors}

As we previously mentioned, to calculate  the form factors
responsible for the rare semileptonic $B_s\to (\eta, \eta')
l^{+}l^{-}$, $(l=e, \mu, \tau)$ and $B_s\to (\eta, \eta') \nu
\bar{\nu}$ decays, we need to calculate the form factors of
$B_s\to \eta_s l^{+}l^{-}/\nu \bar{\nu}$. For this aim,  we start
with the  following three-point  correlation function, which is
constructed from the vacuum expectation value of time ordered
product ${\cal T}$ of interpolating  fields of initial and final
mesons and transition currents, $J^V$ and $J^T$, as follow:
\begin{eqnarray}  \label{eq110}
\Pi_{\mu}^{V,T} = i^2\int
d^{4}xd^4ye^{-ipx}e^{ip^{\prime}y}\langle 0 \vert {\cal T}\left\{
J^s_5(y) J_\mu^{V,T}(0) J^{\dag}_{B_s}(x) \right\} \vert 0
\rangle~,
\end{eqnarray}
where $p$ and $p'$ are initial and final momentums, respectively,
$J_{B_s}(x)= \bar{s}(x)\gamma_{5}b(x)$ and $J
^s_5(y)=\bar{s}(y)\gamma_{5} s(y)$, are the interpolating currents
of the $B_s$ and $\eta_s$ states and
$J_{\mu}^{V}(0)=\bar{s}(0)\gamma_{\mu}b(0) ~$ and $
J_{\mu}^{T}(0)=\bar{s}(0)\sigma_{\mu\nu}q^{\nu }b(0)$ are the
vector and tensor  transition currents  extracted from the
effective Hamiltonian responsible for $B_s\to \eta_s
l^{+}l^{-}/\nu \bar{\nu}$ decays. At quark level, these
transitions are governed by $b\to sl^+l^-$ and  $b\to s\nu
\bar{\nu}$ via penguin and box diagrams (see Fig. (1)).  The
corresponding  effective Hamiltonian is presented in terms of the
Wilson coefficients, $C^{eff}_{7}, C^{eff}_{9}$ and $C_{10}$ as:
\begin{figure}[th]
\begin{center}
\begin{picture}(160,50)
\centerline{ \epsfxsize=13cm \epsfbox{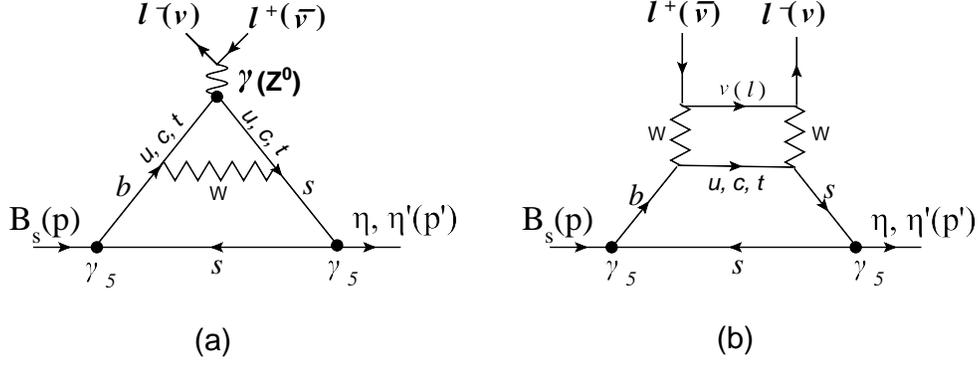}}
\end{picture}
\end{center}
\vspace*{-0.6cm}\caption{Diagrams responsible for the $B_s \to (\eta, \eta')l^{+}l^{-}/\nu \bar{\nu}$
transitions.}\label{F11}
\end{figure}

\begin{eqnarray}  \label{eq111}
\mathcal{H}_{eff} &=& \frac{G_{F}\alpha}{2\sqrt{2} \pi} V_{tb}V_{ts}^\ast %
\Bigg[ C_9^{eff} \, \bar {s} \gamma_\mu (1-\gamma_5) b \, \bar
\ell \gamma_\mu \ell + C_{10}~ \bar {s} \gamma_\mu (1-\gamma_5) b
\, \bar \ell \gamma_\mu \gamma_5
\ell  \nonumber \\
&-& 2 C_7^{eff} \frac{m_b}{q^2}~ \bar {s} ~i\sigma_{\mu\nu} q^\nu
(1+\gamma_5) b \, \bar \ell \gamma_\mu \ell \Bigg]~,
\end{eqnarray}
where $G_{F}$ is the Fermi constant, $\alpha$ is the fine
structure constant at  $Z$ mass scale, and $V_{ij}$ are elements
of the CKM matrix. For $\nu\bar{\nu}$ case, only the term
containing $C_{10}$ is considered. It should be mentioned
that because of the parity conservations, the axial vector and pseudotensor currents
do not contribute to the pseudoscalar--pseudoscalar hadronic
matrix element, i.e.,
\begin{eqnarray}\label{eq115}
\langle P(p')\mid J_{\mu}^{AV}&=&\bar{s}\gamma_{\mu}\gamma_{5}b\mid B_s(p)\rangle
=0~,
\nonumber\\
\langle P(p')\mid J_{\mu}^{PT}&=&\bar{s}~i\sigma_{\mu \nu}q^{\nu}\gamma_{5} b\mid
B_s(p)\rangle =0~,
\end{eqnarray}
where, $P$ stands for $\eta(\eta')$ meson.

From the general aspect of the QCD sum rules, we calculate the
aforementioned correlation function in two different ways. First,
in the hadronic representation, it is calculated in time-like
region in terms of hadronic parameters called phenomenological or
physical side. Second, it is calculated   in space-like region  in
terms of QCD degrees of freedom called the QCD or theoretical
side. The sum rules for the form factors can be obtained  equating
the coefficient of the selected structures from these two
representations of the same correlation function through
dispersion relation and applying double Borel transformation with
respect to the momentums of the initial and final  states to
suppress the contributions coming from the higher states and
continuum.

In order to obtain the phenomenological representation  of the correlation function
given in Eq. (\ref{eq110}), two complete sets of intermediate
states with the same quantum numbers as the interpolating currents   $J_{\eta_s}$
 and $J_{B_s}$ are inserted to sufficient places. As a
result of this procedure, we obtain,
\begin{eqnarray} \label{eq112}
&&\Pi_{\mu}^{V, T}(p^2,p'^2,q^2)= \frac{\langle 0\mid J^s_5 \mid
P(p')\rangle \langle P(p')\mid J_{\mu}^{V, T}\mid B_s(p)\rangle
\langle B_s(p)\mid J^{\dag}_{B_s}\mid
0\rangle}{(p'^2-m_P^2)(p^2-m^2_{B_s})}+\cdots
\end{eqnarray}
where $\cdots$ represents the contributions coming from the higher
states and continuum. The following matrix elements $\langle 0 |
J_{B_s} | P \rangle$ and $\langle 0 | J^s_5 | P \rangle$  are
defined in terms of the leptonic decay constant and four parameters $h^s_P$ as:
\begin{eqnarray}  \label{eq113}
\langle 0| J_{B_s} | B_s \rangle &=&  -i\frac{f_{B_s} m^2_{B_s}}{
m_b+m_s}~,
\nonumber \\
\langle 0| J^s_5|P \rangle &=& -i\frac{h^s_P}{ 2m_s}~.
\end{eqnarray}
where correlating the  $h^s_P$ to $f_s$ and $f_q$, the values
$h^s_{\eta}=-0.053~GeV^3$ and $h^s_{\eta'}=0.065~GeV^3$ are obtained (for details see \cite{FKS}). From
Lorentz invariance and parity considerations, the remaining matrix
element, i.e., transition matrix element in Eq. (\ref{eq112}) is
parameterized in terms of form factors in the following way:
\begin{eqnarray}  \label{eq114}
\langle P(p')\mid J_{\mu}^{V}~\mid B_s(p)\rangle
&=&\mathcal{P}_{\mu}f_{+}(q^{2})+{q}_{\mu}f_{-}(q^{2})~,\nonumber \\
\langle P(p')\mid J_{\mu}^{T}\mid
B_s(p)\rangle &=& \frac{f_T(q^2)}{m_{B_s}+m_P} \Big[
\mathcal{P}_\mu q^2 - q_\mu (m^2_{B_s}-m_P^2) \Big]~,
\end{eqnarray}
where, $f_{+}(q^{2}) , f_{-}(q^{2})$ and $f_{T}(q^{2})$ are the
transition form factors, which only depend on the momentum
transfer squared $q^2$, $ \mathcal{P}_{\mu }=(p+p^{\prime })_{\mu
}$ and $q_{\mu }=(p-p^{\prime })_{\mu }$.

Using Eqs. (\ref{eq113}) and (\ref{eq114})  in Eq. (\ref{eq112}),
we obtain
\begin{eqnarray}  \label{eq116}
\Pi_\mu^{V} (p^2,p^{\prime 2},q^2)&=&\frac{ f_{B_s}
m_{B_s}^2}{2m_s (m_b+m_s)} \frac{h_P^s }%
{ (m_{P}^2 - p^{\prime 2}) (m_{B_s}^2 - p^2)} \Big[ f_+(q^2) \mathcal{%
P}_\mu + f_-(q^2) q_\mu \Big],
\nonumber\\
\Pi_\mu^{T} (p^2,p^{\prime 2},q^2)&=&\frac{ f_{B_s}
m_{B_s}^2}{2m_s (m_b+m_s)} \frac{h_P^s}{%
(m_{P}^2 - p^{\prime 2}) (m_{B_s}^2 - p^2)} \Bigg[ \frac{
f_{T}(q^2)}{(m_{B_s} + m_{P})}
\nonumber\\&\times&~\left(q^2\mathcal{%
P}_\mu -(m_{B_s}^2 - m_{P}^2)q_{\mu}\right)\Bigg].
\end{eqnarray}
For extracting the sum rules for form factors $f_{+}(q^{2})$ and
$f_{-}(q^{2})$, we choose the coefficients of the structures
$\mathcal{P}_{\mu }$  and $q_\mu$ from $\Pi_\mu^{V} (p^2,p^{\prime
2},q^2)$, respectively and the structure $q_\mu$ from $\Pi_\mu^{T}
(p^2,p^{\prime 2},q^2)$ is considered to calculate the form factor
$f_{T}(q^{2})$. Therefore, the correlation functions are written
in terms of the selected structures as:
\begin{eqnarray} \label{eq117}
\Pi_{\mu}^{V}(p^2,p'^2,q^2) \!\!\! &=& \!\!\!
\Pi_+\mathcal{P}_{\mu }  + \Pi_- q_\mu~,\nonumber\\
\Pi_{\mu}^T(p^2,p'^2,q^2) \!\!\! &=& \!\!\! \Pi_{T} q_\mu ~.
\end{eqnarray}

Now, we focus our attention to calculate the  to calculate the QCD
side of the correlation function. This side is calculated at deep
Euclidean space, where  $-p^2\rightarrow\infty$ and
$-p^{'2}\rightarrow\infty$ via operator product expansion (OPE).
For this aim, we write each  $\Pi_{i}$ function (coefficient of
each structure) in terms of the perturbative and non--perturbative
parts as:
\begin{eqnarray} \label{eq118}
\Pi_{i} = \Pi_{i}^{per}
+\Pi_{i}^{non-per}~,
\end{eqnarray}
where $i$ stands for $+$, $-$ and $T$. The perturbative part is written in terms of
double dispersion integral as:
\begin{eqnarray} \label{eq120}\textsl{}
\Pi_{i}^{per} = - \frac{1}{(2 \pi)^2} \int ds^\prime \int ds
\frac{\rho_{i}^{per} (s,s^\prime, q^2)}{(s-p^2) (s^\prime -
p^{\prime 2})}   + \mbox{\rm subtraction terms}~,
\end{eqnarray}
where, the $\rho_{i}^{per} (s,s^\prime, q^2)$ are called spectral
densities. To get the spectral densities, we need to evaluate the
bare loop diagrams in Fig. ( \ref{F11}). Calculating these
diagrams via the usual Feynman integrals with the help of the
Cutkosky rules, i.e.
$\frac{1}{p^2-m^2}\rightarrow-2\pi\delta(p^2-m^2)$, which implies
that all quarks are real, leads to the following spectral
densities:
\begin{eqnarray}\label{eq121}
\rho_+^{per}(s,s',q^2)&=&I_0 N_c \{\Delta +s' -2m_s^2
+2m_bm_s +(E_1+E_2)u\}~,\nonumber\\
\rho_-^{per}(s,s',q^2)&=&I_0 N_c\{%
-\Delta +s' +2m_s^2
-2m_b m_s+(E_1-E_2)u\}~,\nonumber\\
\rho_T^{per}(s,s',q^2)&=&-I_0 N_c\{\Delta (%
m_b-m_s)+s'(m_s-m_b) +2m_s s
+2[m_b(E_1-E_2)\nonumber\\
&&+m_s(E_2-E_1-1)]s'+(E_1-E_2)(m_s-m_b)u\}~,
\end{eqnarray}
where
\begin{eqnarray}\label{eq122}
I_{0}(s,s^{\prime},q^2)&=&\frac{1}{4\lambda^{1/2}(s,s^{\prime},q^2)}
~,
\nonumber \\
\lambda(s,s^{\prime},q^2)&=&s^2+s'^{2}+q^4-2sq^2-2s^{\prime}q^2-2ss^{%
\prime}~,  \nonumber \\
E_{1}&=&\frac{1}{\lambda(s,s^{\prime},q^2)}[2s^{\prime}\Delta-s'{%
}u]~,  \nonumber \\
E_{2}&=&\frac{1}{\lambda(s,s^{\prime},q^2)}[2ss'-\Delta u]~,
\nonumber \\
u&=&s+s^\prime-q^2~,  \nonumber \\
\Delta&=&s+m_s^2-m_b^2~,  \nonumber \\
\end{eqnarray}
and $N_c=3$ is the color factor.

\begin{figure}[th]
\begin{center}
\begin{picture}(160,50)
\centerline{ \epsfxsize=13cm \epsfbox{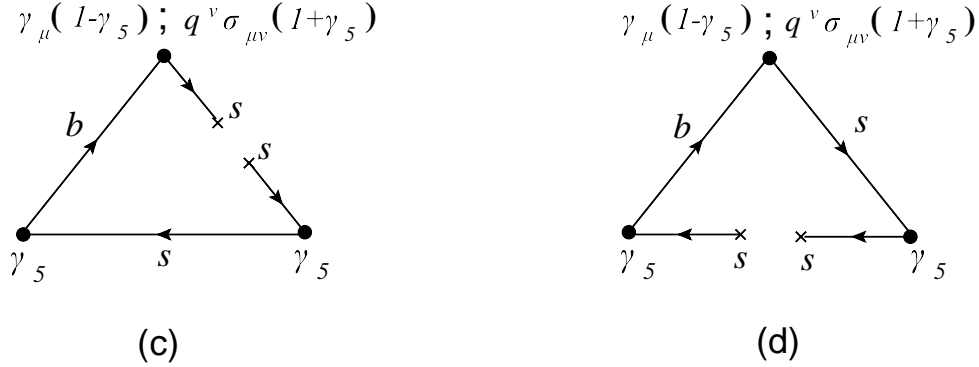}}
\end{picture}
\end{center}
\vspace*{-1cm}\caption{Quark--quark condensate diagrams.}\label{F12}
\end{figure}

\begin{figure}[th]
\begin{center}
\begin{picture}(160,100)
\centerline{ \epsfxsize=13cm \epsfbox{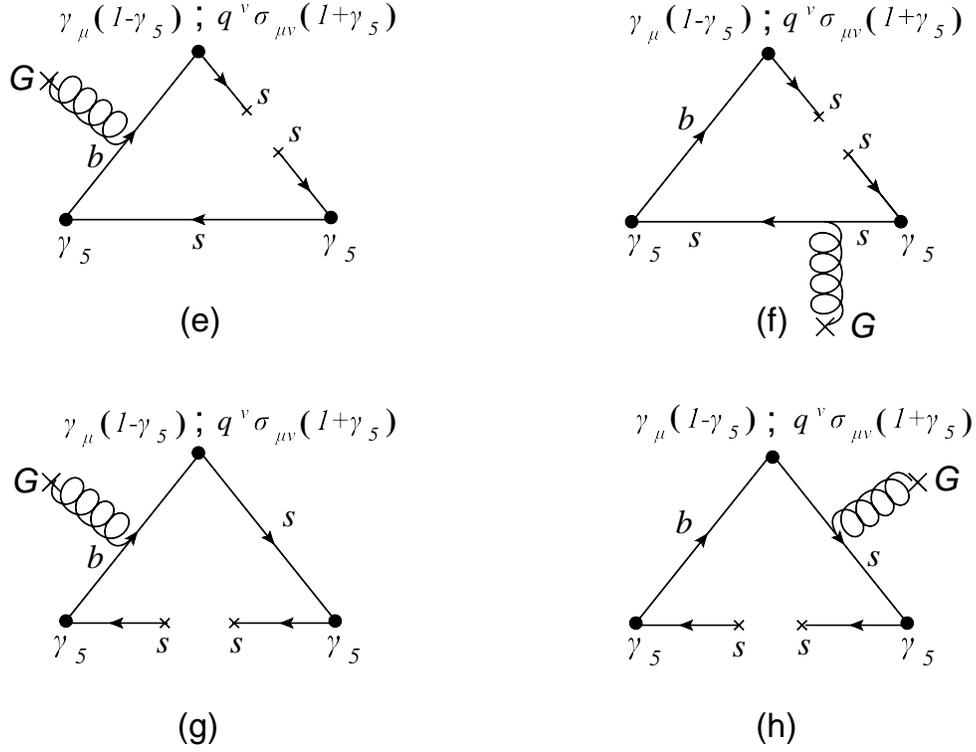}}
\end{picture}
\end{center}
\vspace*{-1cm}\caption{Quark--gluon condensate diagrams.}\label{F13}
\end{figure}

For calculation of  non--perturbative  contributions in QCD side,
the condensate terms of OPE are considered. The condensate term of
dimension $3$ is related to contribution of quark condensate. Fig
.(\ref{F12}) shows quark--quark condensate diagrams of dimension
$3$. It should be reminded that the quark condensate are
considered only for  light quarks and the heavy quark condensate
is suppressed by inverse powers of the heavy quark mass.
The contribution of the  diagram (c) in  Fig .(\ref{F12}) is zero
since applying double Borel transformation with respect to the
both variables $p^2$ and ${p^{'}}^2$ kills its contribution,
because only one variable appears in the denominator in this case.
Therefore as dimension $3$, we consider only  diagram (d) in Fig
.(\ref{F12}). The  dimension $4$ operator in OPE is the
gluon--gluon condensate. Our calculations show that in this case,
the gluon--gluon condensate contributions are very small in
comparison with the quark--quark and quark-gluon condensates
contributions and we can easily ignore their contributions. The
next operator is dimension $5$  quark--gluon condensate. The
diagrams corresponding to quark--gluon condensate are presented in
Fig. (\ref{F13}).
Contributions of the  diagrams (e) and (f)  vanish  with the same
reason as  for diagram (c) in Fig .(\ref{F12}). Therefore, only
diagrams (g) and (h) contribute to the  non--perturbative part of
dimension $5$. In QCD sum rule approach, the  OPE is truncated at
some finite order such that Borel transformations play an
important role in this cutting. Mainly, the proper regions of the
Borel parameters are adopted by demanding that in the truncated
OPE, the condensate term with the highest dimension constitutes a
small fraction of the total dispersion integral. In  the next
section, we will explain how these proper regions are obtained.
Hence, we will not consider the condensates with $d \geq 6$  that
play a minor role in our calculations.

The explicit expressions of $\Pi _{i}^{non-per}$,
 are given in the Appendix--A.

The next step is to apply the double Borel transformations with respect
to the $p^2 (p^2\to M_1^2)$ and $p'^2 (p'^2\to M_2^2)$ on the
phenomenological as well as the perturbative and non--perturbative
parts of the QCD side and equate the two
representations. As a result, the following sum rules for
the form factors are
derived:
\begin{eqnarray}\label{eq123}
f'_{i}(q^2) \!\!\! &=& \!\!\! \frac{(m_b+m_s) (2m_s)}{ f_{B_s}
m_{B_s}^2 h_P^s}
e^{m_{B_s}^2/M_1^2} e^{m_{P}^2/M_2^2}  \nonumber \\
&\times& \!\!\!\Bigg\{-\frac{1}{4 \pi^2}  \int_{2m_s^2}^{s'_0} ds'
\int_{s_L}^{s_0}ds \rho_{i}^{per} (s,s',q^2) e^{-s/M_1^2}
e^{-s'/M_2^2} +  \tilde{B}\Pi_{i}^{non-per}(p^2,p'^2,q^2)%
\Bigg\}~,\nonumber \\
\end{eqnarray}
where, $f'_+(q^2)=f_+(q^2)$, $f'_-(q^2)=f_-(q^2)$ and
$f'_T(q^2)=-f_T(q^2)(m_{B_s}-m_P)$.  The $s_0$ and $s'_0$ are the
continuum thresholds in initial and final channels, respectively
and $s_L$ is the lower limit of the integral over $s$. It is
obtained  as:
\begin{eqnarray}\label{eq124}
s_{L}=\frac{(m_s^2+q^2-m_b^2-s^{\prime})(m_b^2 s^{\prime}-q^2
m_s^2)}{(m_b^2-q^2)(m_s^2-s^{\prime})}~.
\end{eqnarray}
Also the operator $\tilde{B}$ in Eq. (\ref{eq123}) is defined as:
\begin{equation}\label{eq125}
\tilde{B}={\cal{B}}_{p^2}(M_1^2){\cal{B}}_{{p^{'}}^2}(M_2^2)~,
\end{equation}
where, $M^2_1$ and  $M^2_2$ are Borel mass parameters.
It should be also noted that to subtract the contributions of the
higher states and the continuum the quark--hadron duality
assumption is also used,
\begin{eqnarray}\label{eq127}
\rho^{higher states}(s,s') = \rho^{OPE}(s,s') \theta(s-s_0)
\theta(s'-s'_0)~.
\end{eqnarray}

\section{Numerical analysis}
We are now ready to present our numerical analysis of the form factors $%
f_{+}(q^2),~f_{-}(q^2)$ and $f_{T}(q^2)$ and calculate  branching fractions and
longitudinal lepton polarization asymmetries. In
our numerical calculations, we  use the following values for input
parameters: $m_s=0.13~GeV$,
$m_b=4.8~GeV$, $m_{\eta}=(547.51\pm 0.18)~MeV$,
$m_{\eta'}=(957.78\pm 0.14)~MeV$, $m_{B_s}=(5366.3\pm 0.6)~MeV$ \cite{PDG}, $|
V_{tb}V_{ts}^*|=0.0385$,
$C^{eff}_{7}=-0.313$, $C_{9}=4.344$,
$C_{10}=-4.669$ \cite{Buras}, $f_{B_s}=(209\pm38)~MeV$ \cite{Rolf},
$m_0^2=(0.8\pm 0.2)~GeV^2$, $\langle s\bar{s}\rangle=(0.8\pm 0.2)
\langle u\bar{u}\rangle$ and $\langle u\bar{u}\rangle=-(0.240\pm
0.010)^3~GeV^3$.

The sum rules for the form factors contain also four auxiliary
parameters, namely Borel mass squares, $M_1^2$ and $M_2^2$ and
continuum thresholds, $s_0$ and $s_0$. These are not physical
quantities, so our results should be independent of them. The
parameters $s_0$ and $s_0^\prime$ are not totally arbitrary but
they are related to the energy of the first excited stateS with the
same quantum numbers as the interpolating currents. They are determined from the conditions that
guarantee the sum rules to have the best stability in the allowed
$M_1^2$ and $M_2^2$ regions. The value of continuum threshold
$s_0$ calculated from the two--point QCD sum rules are taken to be
$s_0=(34.2\pm2)~GeV^2$ \cite{Ball}. We use also the range,
$(m_P+0.3)^2\leq s'_0\leq(m_P+0.5)^2~GeV^2$ in $P=\eta(\eta')$
channel. The working regions for $M_1^2$ and $M_2^2$ are
determined demanding that not only the contributions of the higher
states and continuum are effectively suppressed, but contributions
of the higher dimensional operators are also small. Both
conditions are satisfied in the  regions, $12~GeV^2 \le M_1^2 \le
22~GeV^2$ and $4~GeV^2 \le M_2^2 \le 10~GeV^2$.

\begin{figure}[th]
\vspace*{4.cm}
\begin{center}
\begin{picture}(160,20)
\put(0,-20){ \epsfxsize=8cm \epsfbox{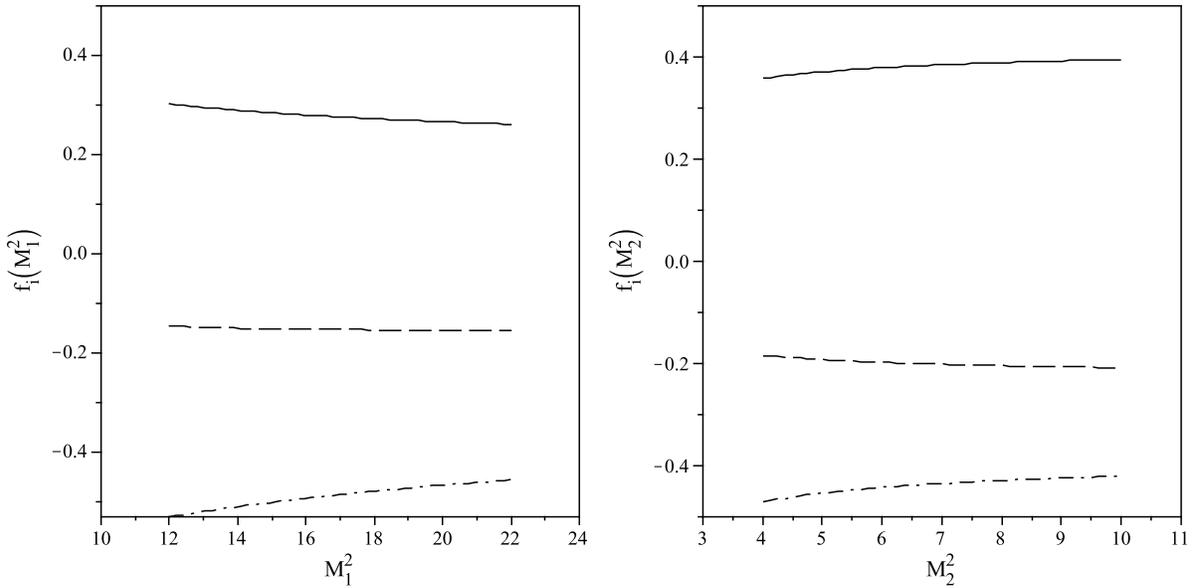}\epsfxsize=8cm
\epsfbox{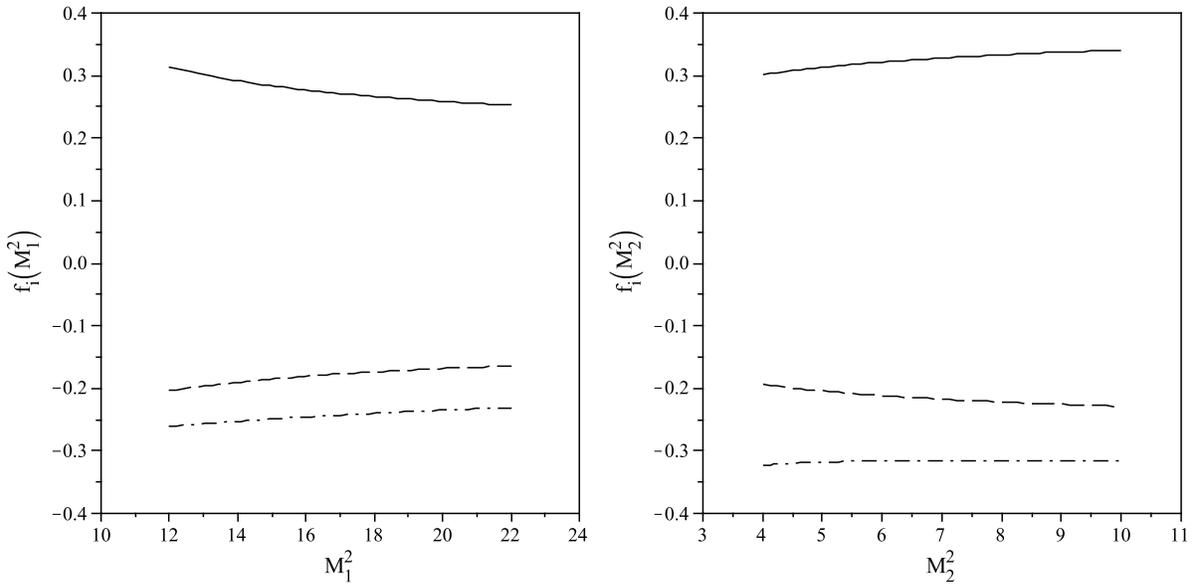} }
\end{picture}
\end{center}
\vspace*{1cm} \caption{The dependence of the form factors on
$M_1^2$ and $M_2^2$ for $B_s\to \eta_s$ decay  when
$m_P=m_{\eta}$. The solid,  dashed and dashed-dotted lines
correspond to the $f_+$,  $f_-$ and $f_T$,
respectively.}\label{F21}
\end{figure}

\begin{figure}[th]
\vspace*{4.cm}
\begin{center}
\begin{picture}(160,20)
\put(0,-20){ \epsfxsize=8cm \epsfbox{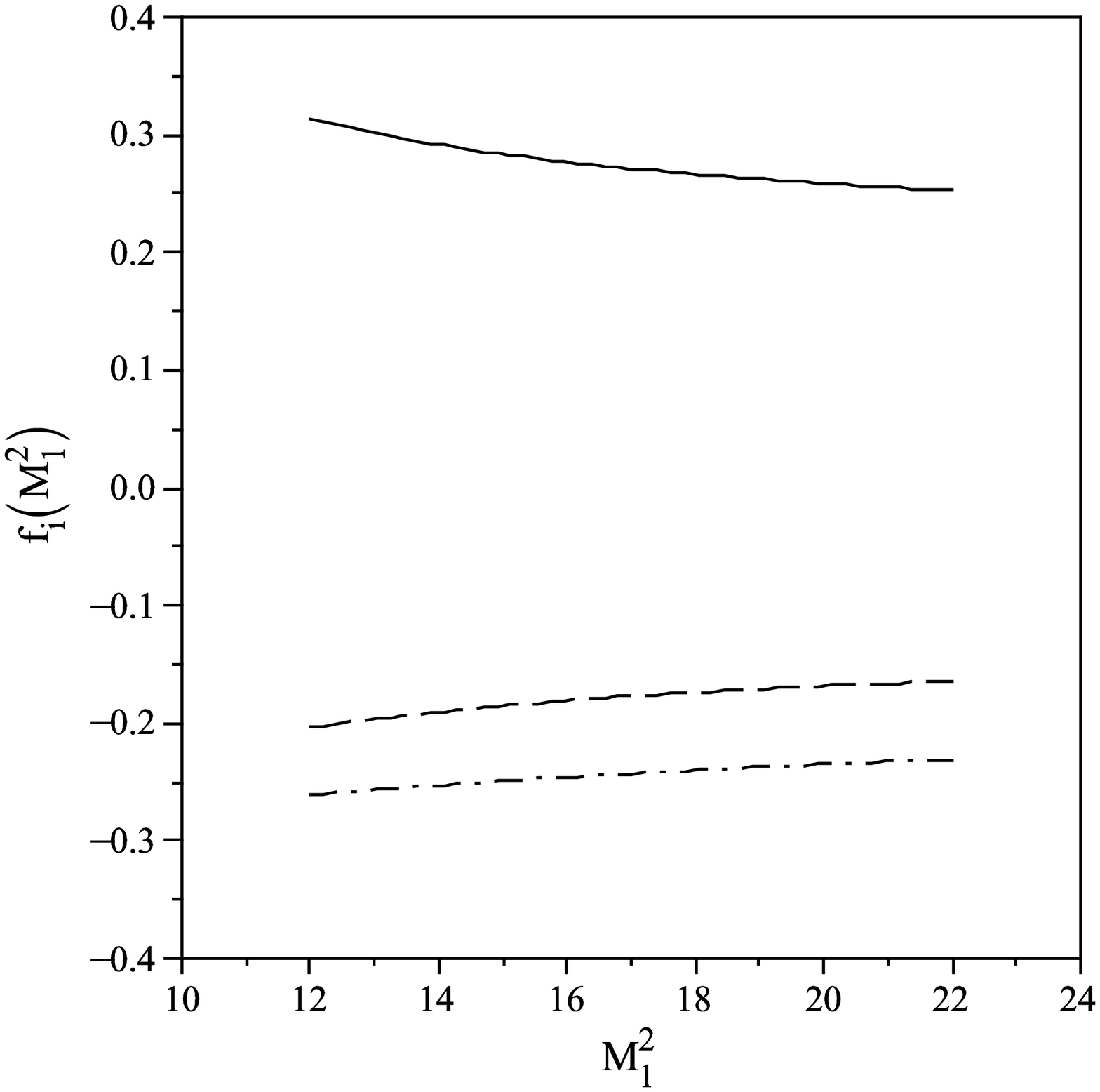}\epsfxsize=8cm
\epsfbox{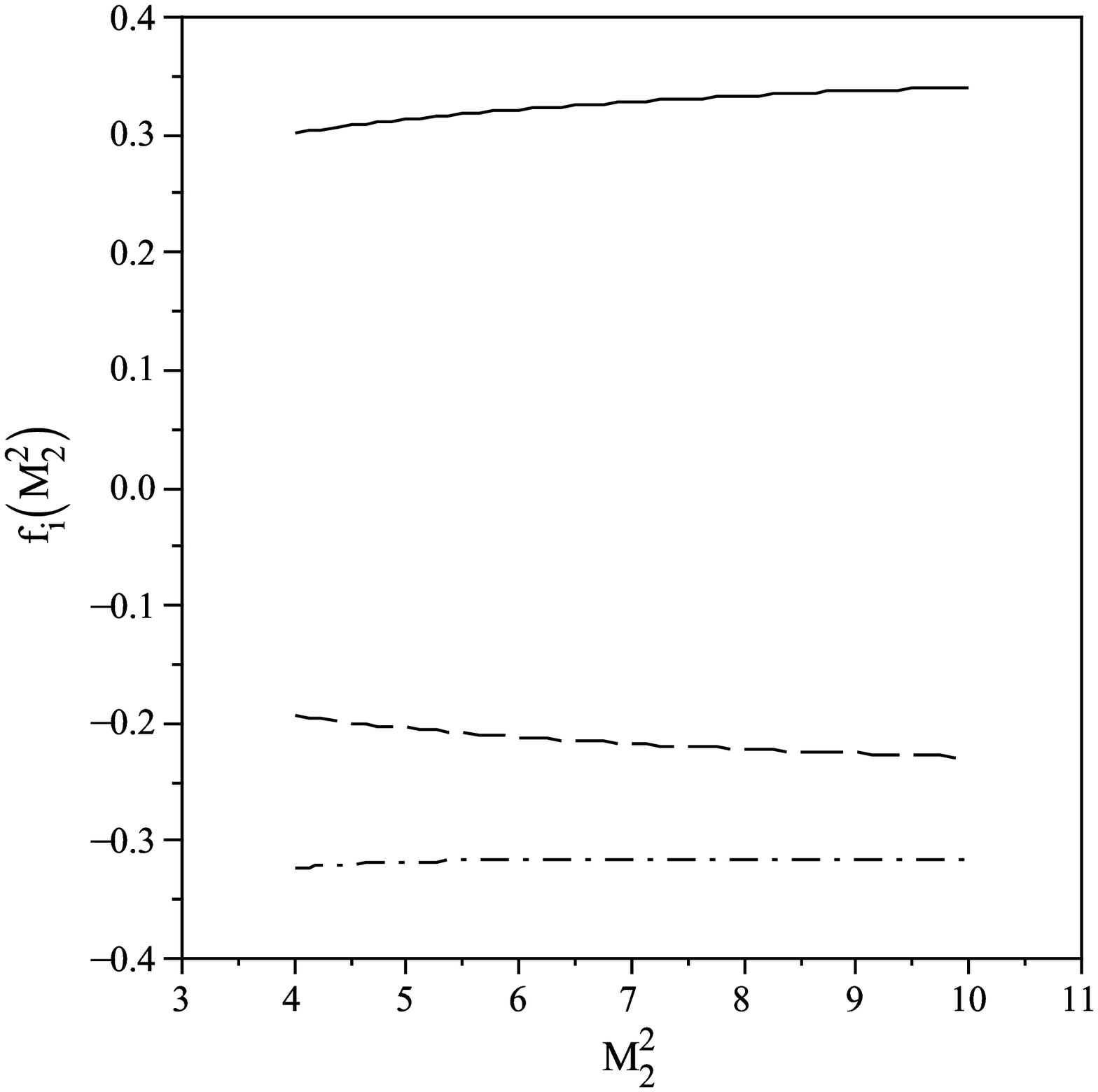} }
\end{picture}
\end{center}
\vspace*{1cm} \caption{The dependence of the form factors on
$M_1^2$ and $M_2^2$ for $B_s\to \eta_s$ decay  when
$m_P=m_{\eta'}$. The solid,  dashed and dashed-dotted lines
correspond to the $f_+$,  $f_-$ and $f_T$,
respectively.}\label{F22}
\end{figure}

The dependence of the form factors  $f_{+},~f_{-}$ and $f_{T}$ on
$M_1^2$ and $M_2^2$ for $B_s \rightarrow \eta_s$  transition when
$m_P=m_{\eta}$ are shown in Fig. \ref{F21}. The Fig. \ref{F22}, also
depicts the dependence of the same form factors on Borel mass
parameters for $B_s \rightarrow \eta_s$ decay when
$m_P=m_{\eta'}$.
These figures show a good stability of the form factors with
respect to the Borel mass parameters in the working regions. Using
these regions for $M_1^2$ and $M_2^2$, our numerical analysis
shows that the contribution of the non--perturbative part to the QCD side  is about
$21\% $ of the total and the main contribution comes from the
perturbative part.

Now, we proceed to present the $q^2$ dependency of the form
factors. Since the form factors $f_{\pm}(q^2)$ and $f_{T}(q^2)$
are calculated in the space-like ($q^2<0$) region,  we should
analytically continue them to the time-like ($q^{2}>0$) or
physical region. Hence, we should  change $q^2$ to $-q^2$. As we
previously mentioned, the form factors are truncated at
approximately, $1~GeV$ below the perturbative cut. Therefore, to
extend our results to the full physical region, we look for
parametrization of the form factors in such a way that in the
reliable  region the results of the parametrization coincide with
the sum rules predictions. Our numerical calculations show that
the sufficient parametrization of the form factors with respect to
$q^2$ is:
\begin{equation}  \label{eq21}
f_{i}(q^2)=\frac{f_{i}(0)}{1+ \alpha\hat{q}+ \beta\hat{q}%
^2}~,
\end{equation}
where $\hat{q}=q^2/m_{B_s}^2$.
The values of the parameters $%
f_{i}(0), ~\alpha$ and  $\beta$ are given in the Table \ref{T21}
taking $M_{1}^2=12~GeV^2$ and $ M_{2}^2=5~GeV^2$. This Table also
contains the predictions of the light-front quark model (LFQM).
\begin{table}[th]
\centering
\begin{tabular}{ccc|ccc}
\hline\hline
                               &$B_s\to \eta_s(P=\eta) $&                &
                               &$B_s\to \eta_s(P=\eta')$&                \\
\hline
                    Parameters &$\quad$ $\quad$This work$\quad$ $\quad$ & LFQM\cite{HMC} &
$\quad$     Parameters &$\quad$ $\quad$This work$\quad$ $\quad$ & LFQM\cite{HMC} \\
\hline
                  $f_{+}(0)$   & $0.364 \pm 0.120$           &$0.291$&
$\quad$           $f_{+}(0)$   & $0.337 \pm 0.111$           &$0.291$ \\
                  $\alpha$     & $-0.333\pm 0.107$           &$-1.574$&
$\quad$           $\alpha$     & $-0.495\pm 0.158$           &$-1.575$ \\
                  $\beta$      & $-0.694\pm 0.222$           &$0.751$&
$\quad$           $\beta$      & $-0.820\pm 0.262$           & $0.770$  \\
\hline
                  $f_{-}(0)$   & $-0.189\pm 0.062$           &$-0.231$&
$\quad$           $f_{-}(0)$   & $-0.193\pm 0.064$           &$-0.225$ \\
                  $\alpha$     & $-0.833\pm 0.267$           &$-1.582$&
$\quad$           $\alpha$     & $-1.028\pm 0.329$           &$-1.570$ \\
                  $\beta$      & $-0.168\pm 0.054$           &$0.825$ &
$\quad$           $\beta$      & $-0.048\pm 0.015$           &$0.835$  \\
\hline
                  $f_{T}(0)$   & $-0.444\pm 0.147$           &$-0.280$&
$\quad$           $f_{T}(0)$   & $-0.424\pm 0.140$           &$-0.300$ \\
                  $\alpha$     & $-0.453\pm 0.145$           &$-1.561$&
$\quad$           $\alpha$     & $-0.596\pm 0.191$           &$-1.561$ \\
                  $\beta$      & $-0.355\pm 0.114$           &$0.782$ &
$\quad$           $\beta$      & $-0.381\pm 0.122$           &$0.802$  \\
\hline\hline
\end{tabular}
\vspace{0.4cm} \caption{Parameters appearing in the fit function
for form factors of $B_s \to \eta_s$ in two
approaches.}\label{T21}
\end{table}

\begin{table}[h]
\centering
\begin{tabular}{cccccc}

\hline\hline      Mode     $\quad$& Form factors$\quad$&$\quad$ This work  $\quad$&  LFQM\cite{HMC} &$\quad$LFQM\cite{GL} &  CQM \cite{GL}  \\
\hline
                                  &$f_+(0)$ &$0.364 \pm0.120$ &0.291            & 0.354          & 0.357           \\
$B_s \to \eta_s(P=\eta)$          &$f_-(0)$ &$-0.189\pm0.062$ &-0.231           & -0.360         & -0.304          \\
                                  &$f_T(0)$ &$-0.444\pm0.147$ &-0.280           & -0.369         & -0.365          \\
\hline
                                  &$f_+(0)$ &$0.337 \pm0.111$ &0.291            & 0.354          & 0.357           \\
$B_s \to \eta_s(P=\eta')$         &$f_-(0)$ &$-0.193\pm0.064$ &-0.225           &-0.324          & -0.304          \\
                                  &$f_T(0)$ &$-0.424\pm0.140$ &-0.300           & -0.404         & -0.390          \\

\hline\hline
 \end{tabular}
\vspace{0.8cm} \caption{The form factors of the $B_s\to \eta_s$
decay for $M_{1}^2=12~GeV^2$ and  $ M_{2}^2=5~GeV^2$ at $q^2=0$ in
different approaches: this work (3PSR), light-front quark model
(LFQM) and constituent quark model (CQM).}\label{T210}
\end{table}
The values of the form factors at $q^2=0$ are also compared with the predictions of the other nonperturbative approaches such as, LFQM and constituent quark model (CQM)
 in Table \ref{T210}.
The dependence of the form factors $f_{+}(q^2)$, $f_{-}(q^2)$ and
$f_{T}(q^2)$ on $q^2$ extracted from the fit function are given in
Figs. (\ref{F23}) and (\ref{F24}) for the $P= \eta$ and $P= \eta'$
cases, respectively. These figures also contain the values of form
factors obtained directly from our sum rules in reliable region.
These values coincide well with the values obtained from the fit
function below the perturbative cut. Therefore, the aforementioned
fit parametrization better describe our form factors. The form
factors of $B_s \to \eta$ and $B_s \to \eta'$ are obtained using
values in Table \ref{T21} and also Eq. (\ref{Fee}).

Now, we would like to evaluate  the branching ratios
for the considered decays. Using the parametrization of these
transitions in terms of the form factors, we get \cite{Chen}:
\begin{eqnarray}\label{eq128}
\frac{d{\Gamma}}{dq^2}(B_s\rightarrow P {\nu} \bar \nu) &=& \frac{%
A~G_{_F}^2{\mid}V_{ts}V^*_{tb}{\mid}^2
m_{B_s}^3\alpha^2} {2^8 \pi ^5}~\frac{{\mid}D_{\nu}(x_t){\mid}^2}{sin^4\theta_W}~\phi%
^{3/2}(1,\hat{r},\hat{s}){\mid}f_+ (q^2){\mid}%
^2~,\nonumber\\
\frac{d\Gamma  }{dq^2}\left( B_s \rightarrow P l^+l^-\right) &=&
\frac{A~G_{_F}^2{\mid}V_{ts}V^*_{tb}{\mid}^2 m_{B_s}^3\alpha^2}
{3\cdot 2^9\pi ^5}v \phi^{1/2}(1,\hat{r},\hat{s})\left[
\left(1+\frac{2\hat{l}} {\hat{s}}\right)
\phi(1,\hat{r},\hat{s})\alpha _1+12~\hat{l}\beta_1\right]~,\nonumber\\
\end{eqnarray}
where $A=\sin^2\varphi$ for $B_s \rightarrow \eta$ and
$A=\cos^2\varphi$ for $B_s \rightarrow \eta'$ transitions. The
$\hat{r}$, $\hat{s}$, $\hat{l}$, $x_t$ and ${\hat m}_b$ and the
functions $v$, $\phi(1,\hat{r},\hat{s})$, $D_{\nu}(x_t)$,
$\alpha_1$ and $\beta_1$ are defined as:
\begin{eqnarray}\label{eq129}
\hat{r} &=& \frac{m_{P}^2}{m_{B_s}^2}~,\quad \hat{s}
=\frac{q^2}{m_{B_s}^2}~,\quad \hat{l} = \frac{m_{l}^2}{
m_{B_s}^2}~,\quad x_t =
\frac{m_t^2}{m_W^2}~,\quad {\hat m}_b = \frac{m_b}{m_{B_s}}~,\nonumber\\
v &=&\sqrt{1-\frac{4\hat{l}}{\hat{s}}}~,\nonumber\\
{\phi}(1,\hat{r},\hat{s}) &=& 1+\hat{r}^2
+\hat{s}^2-2\hat{r}-2\hat{s}-2\hat{r}\hat{s}~,\nonumber\\
D_{\nu}(x_t)&=&\frac{x_t}{8}\Bigg(\frac{2+x_t}{x_t-1}+\frac{3x_t-6}{(x_t-1)^2}\ln x_t\Bigg)~,\nonumber\\
\alpha_1 &=& \biggl|C_{9}^{\rm eff}\,f_{+}(q^2) +\frac{2\,{\hat
m}_b\,C_{7}^{\rm eff}\,f_{T}(q^2)}{1+\sqrt{\hat{r}}}\biggr|^{2}
+|C_{10}f_{+}(q^2)|^{2} ~,\nonumber\\
\beta_1 &=& |C_{10}|^{2}\biggl[ \biggl( 1+\hat{r}-{\hat{s}\over
2}\biggr) |f_{+}(q^2)|^{2}+\biggl( 1-\hat{r}\biggr) {\rm
Re}(f_{+}(q^2)f_{-}^{*}(q^2))+\frac{1}{2}\hat{s}|f_{-}(q^2)|^{2}\biggr].~
\end{eqnarray}

Integrating Eq.$~$(\ref{eq128}) over $q^2$ in the whole physical
region and using the total mean lifetime $\tau_{B_{s}}=
(1.466\pm0.059)~ps$ \cite{PDG}, the branching ratios of the $B_s
\rightarrow (\eta, \eta')l^{+}l^{-}/\nu\bar{\nu}$ are obtained as
presented in Table \ref{T22}.
\begin{table}[th]
\centering
\begin{tabular}{c|ccccccc}
\hline \hline
Mode&This work&LFQM\cite{HMC}&LFQM\cite{GL}&CQM\cite{GL}&set A\cite{CCD}&set B\cite{CCD}&set C\cite{CCD}\\
\hline
$Br(B_{s} \to \eta \nu\bar{\nu})\times 10^{6}$ & $1.35\pm0.56 $ & $1.54$ & $2.56(2.34)$ & $2.38(2.17)$ & $0.95\pm0.2 $ & $2.2\pm0.7$&$2.9\pm1.5$\\
$Br(B_{s} \to \eta' \nu\bar{\nu})\times 10^{6}$& $1.33\pm0.55 $ & $1.47$ & $2.36(2.52)$ & $2.23(2.38)$ & $0.9\pm0.2  $ &$1.9\pm0.5  $&$2.4\pm1.3$\\
$Br(B_{s} \to \eta \mu^+ \mu^-)\times 10^{7}$  & $2.30\pm0.97 $ & $2.09$ & $3.75(3.42)$ & $3.42(3.12)$ & $1.2\pm0.3  $ &$2.6\pm0.7  $&$3.4\pm1.8$\\
$Br(B_{s} \to \eta'\mu^+ \mu^-)\times 10^{7}$  & $2.24\pm0.94 $ & $1.98$ & $3.40(3.63)$ & $3.19(3.41)$ & $1.1\pm0.3  $ &$2.2\pm0.6  $&$2.8\pm1.5$\\
$Br(B_{s} \to \eta \tau^+ \tau^-)\times 10^{8}$& $3.73\pm1.56 $ & $5.14$ & $7.33(6.70)$ & $7.33(6.70)$ & $3  \pm0.5  $ &$8   \pm1.5 $&$10 \pm5.5$\\
$Br(B_{s} \to \eta'\tau^+ \tau^-)\times 10^{8}$& $2.80\pm1.18 $ & $2.86$ & $4.66(5.00)$ & $4.04(4.30)$ & $1.55\pm0.3 $ &$3.85\pm0.75$&$4.7\pm2.5$\\
\hline\hline
\end{tabular}
\vspace{0.4cm} \caption{The branching ratios in different models
corresponding to $\varphi=41.5^\circ$. The values in parentheses
related to $\varphi=39.3^\circ$.} \label{T22}
\end{table}
In this Table, we show only the values obtained  considering
the short distance (SD) effects contributing to the Wilson
coefficient $C_9^{\rm eff}$ for charged lepton case. The effective
Wilson coefficient $C^{\rm eff}_9$ including both the SD and long distance (LD) effects is~\cite{Buras}:
\begin{eqnarray}\label{eq23}
C^{\rm eff}_9(s) = C_9 + Y_{SD}(s) + Y_{LD}(s).
\end{eqnarray}
The LD  effect contributions are due to the $J/\psi$ family. The explicit expressions of the $Y_{SD}(s)$ and $Y_{LD}(s)$
can be found in \cite{Buras} (see also \cite{Faessler}).
Table \ref{T22} also includes a comparison between our results and
predictions of the other approaches including the LFQM,
CQM and other methods \cite{CCD}.
Note that, the results presented as \cite{CCD} are not the results
directly obtained by analysis of the $B_s \to \eta(\eta')$, but
they have been found relating the form factors of $B_s \to \eta_s$
to the form factors of $B\to K$ using the quark flavor scheme (see
\cite{CCD}). Hence, the comparison of our results with the predictions of \cite{CCD} is an approximate  and for the exact comparison, the form factors should be directly available. In this Table, the set A refers to the values
computed using short-distance QCD sum rules, set B shows the
results obtained  by light-cone QCD sum rules and set C
corresponds to the results calculated via light-cone QCD sum rules
within the Soft Collinear Effective Theory (SCET). From Table
\ref{T22}, we see a good consistency in order of magnitude between our results and
predictions of the other non-perturbative approaches. Here, we
should also stress that the results obtained for the electron are
very close to the results of the muon and for this reason, we only
present the branching ratios for muon in our Tables.

In this part, we would like to  present the branching ratios
including LD effects. We introduce some cuts
around the resonances of $J/\psi$ and $\psi^{\prime }$ and study
the following three regions for muon:
\begin{eqnarray}\label{eq26}
\mbox{I}: &&\ \ \ \ \ \ \ \ \sqrt{q^2_{min}} \;\leq\; \sqrt{q^2}
\;\leq\; M_{J/\psi }-0.20,
\nonumber\\
\mbox{II}: && M_{J/\psi}+0.04 \;\leq\; \sqrt{q^2} \;\leq\;
M_{\psi^{\prime}}-0.10,
\nonumber \\
\mbox{III}: &&\ \ M_{\psi^{\prime}}+0.02 \;\leq\; \sqrt{q^2}
\;\leq\; m_{B_s}-m_P.
\end{eqnarray}
and for tau:
\begin{eqnarray}\label{eq27}
\mbox{I}: & \ \ \ \ \ \ \ \sqrt{q^2_{min}} \;\leq\; \sqrt{q^2}
\;\leq\; M_{\psi'} - 0.02,
\nonumber\\
\mbox{II}: & M_{\psi'} + 0.02\; \leq\; \sqrt{ q^2}\; \leq\;
m_{B_s}-m_P.
\end{eqnarray}
where $\sqrt{q^2_{min}}=2m_l$. In Tables \ref{T23} and \ref{T24},
we present the branching ratios for muon and tau obtained using
the regions shown in Eqs. (\ref{eq26}) and (\ref{eq27}),
respectively. The errors presented in Tables \ref{T22}, \ref{T23} and \ref{T24} are due to uncertainties in determination of the auxiliary parameters, errors in input parameters, systematic errors in QCD sum rules 
as well as the errors associated to the following approximations used in the present work: a) the form factors are calculated in the low $q^2$ region and  extrapolated to high $q^2$ using the fit parametrization
 in Eq. (\ref{eq21}), b) the  hadronic operators in the  considered Hamiltonian can   receive sizable non-factorizable
       corrections and  the corresponding matrix elements may also be sensitive to the isosinglet content of the  $\eta$ and $\eta'$ mesons. We  show the dependency of the differential
branching ratios on  $q^2$ (with and without  LD effects for
charged lepton case) in Figs. (\ref{F25})-(\ref{F210}).
\begin{table}[th]
\centering
\begin{tabular}{cccc}
\hline\hline
Mode     &     $\mbox{I}$      &      $\mbox{II}$      &      $\mbox{III}$\\
\hline
$Br(B_s\to \eta    \mu^+\mu^-)$~~ &~~$(1.76\pm0.72)\times 10^{-7}$~~&~~$(2.20\pm0.90)\times 10^{-8} $~~&~~$(2.28\pm0.93)\times 10^{-8}$\\
$Br(B_s\to \eta'   \mu^+\mu^-)$~~ &~~$(1.81\pm0.74)\times 10^{-7}$~~&~~$(2.24\pm0.92)\times 10^{-8} $~~&~~$(1.32\pm0.54)\times 10^{-8}$\\
\hline\hline
\end{tabular}
\vspace{0.01cm} \caption{The branching ratios of the semileptonic
$B_s\to (\eta, \eta')\mu^+\mu^-$  decays including  LD
effects.}\label{T23}
\end{table}
\begin{table}[th]
\centering
\begin{tabular}{ccc}
\hline\hline
Mode     &     $\mbox{I}$      &      $\mbox{II}$      \\
\hline
$Br(B_s\to \eta \tau^+\tau^-)  $~~ &~~$(0.40\pm0.16)\times 10^{-9}$~~&~~$(3.16\pm1.26)\times 10^{-8}$\\
$Br(B_s\to \eta'  \tau^+\tau^-)$~~ &~~$(0.43\pm0.17)\times 10^{-9}$~~&~~$(2.27\pm0.90)\times 10^{-8}$\\
\hline\hline
\end{tabular}
\vspace{0.01cm} \caption{The branching ratios of the semileptonic
$B_s\to (\eta, \eta')\tau^+\tau^-$  decays including LD
effects.}\label{T24}
\end{table}

Finally, we want to calculate  the longitudinal lepton polarization
asymmetry for considered decays. It is given as \cite{Chen}:
\begin{eqnarray}
P_L=\frac{2v}{(1+\frac{2\hat{l}}{\hat{s}})\phi(1,\hat{r},\hat{s})\alpha_1+12
\hat{l}\beta_1}{\rm{Re}}\left[\phi(1,\hat{r},\hat{s})\left(C_9^{eff}f_+(q^2)-\frac{2C_7
f_T(q^2)}{1+\sqrt{\hat{r}}}
\right)(C_{10}f_+(q^2))^*\right],\nonumber \\
\end{eqnarray}
where $v, ~\hat{l}, ~\hat{r}, ~\hat{s}, ~\phi(1,\hat{r},\hat{s}),
\alpha_1$ and $\beta_1$ were defined before. The dependence of the
longitudinal lepton polarization asymmetries for the $B_s\to
(\eta, \eta')l^+l^-$ decays on the transferred momentum square
$q^2$ with and without LD effects are plotted in Figs. \ref{F211}
and \ref{F212}.

As a result, the order of the obtained values for branching ratios
as well as the  longitudinal lepton polarization asymmetries show
a possibility to study the considered transitions at LHC. Any
experimental measurements on the presented quantities and those
comparisons with the obtained results can give valuable
information about the nature of the $\eta$ and $\eta'$ mesons and
strong interactions inside them.

\section*{Acknowledgments}
Partial support of Shiraz university research council is
appreciated.

\clearpage

\appendix
\begin{center}
{\Large \textbf{Appendix--A}}
\end{center}

\setcounter{equation}{0} \renewcommand{\theequation}

In this appendix, the explicit expressions of the
$\Pi_{i}^{non-per}$ are given,
\begin{eqnarray*}
\Pi _{+}^{non-per}(p^2,p'^2,q^2) &=&\langle
s\bar{s}\rangle\Bigg(-\frac{m_s}{2rr'}
+\frac{4m_0^2m_s-2m_0^2m_b+3m_s^3-3m_s^2m_b}{12rr^{'2}}\\
&&+\frac{m_0^2m_b^3-m_0^2m_s^3+3m_s^4-3m_b^3m_s^2-2m_0^2m_b^2m_s+2m_0^2m_bm_s^2}{12r^2r^{'2}}\\
&&+\frac{m_0^2m_sq^2-m_0^2m_bq^2+3m_b^2m_s^3-3m_bm_s^3+3m_bm_s^2q^2-3m_s^3q^2}{12r^2r^{'2}}\\
&&+\frac{2m_0^2m_s-4m_0^2m_b+3m_sm_b^2-3m_s^2m_b}{12r^2r'}\\
&&+\frac{m_0^2m_b^3-2m_b^3m_s^2+2m_b^2m_s^3-m_0^2m_b^2m_s}{4r^3r'}\\
&&+\frac{2m_s^5-m_0^2m_s^3-2m_bm_s^4+m_0^2m_bm_s^2}{4rr^{'3}}\Bigg),
\end{eqnarray*}
\begin{eqnarray*}
\Pi _{-}^{non-per}(p^2,p'^2,q^2) &=&\langle s\bar{s}\rangle \Bigg(
\frac{2m_0^2m_b-9m_s^3+3m_s^2m_b}{12rr^{'2}}\\
&&+\frac{3m_s^5-m_0^2m_b^3-m_0^2m_s^3+3m_b^3m_s^2+m_0^2m_bq^2+12m_0^2m_sq^2}{12r^2r^{'2}}\\
&&+\frac{3m_b^2m_s^3+3m_bm_s^4-3m_bm_s^2q^2-3m_s^3q^2}{12r^2r^{'2}}\\
&&+\frac{2m_0^2m_s-3m_sm_b^2+6m_s^2m_b-3m_s^2m_b}{12r^2r'}\\
&&+\frac{2m_b^3m_s^2-m_0^2m_b^3+2m_b^2m_s^3-m_0^2m_b^2m_s}{4r^3r'}\\
&&+\frac{2m_s^5-m_0^2m_s^3+2m_bm_s^4-m_0^2m_bm_s^2}{4rr^{'3}}\Bigg),
\end{eqnarray*}
\begin{eqnarray*}
\Pi _{T}^{non-per}(p^2,p'^2,q^2)
&=&\langle s\bar{s}\rangle\Bigg(\frac{2m_s^3+2m_sm_b+m_0^2}{4rr'}\\
&&+\frac{3m_s^5-m_0^2m_b^2+3m_s^2q^2-m_0^2q^2+3m_s^4-6m_s^2m_b^2-m_sm_0^2m_b}{6rr^{'2}}\\
&&+\frac{m_0^2m_b^4-m_0^2m_s^4+3m_s^6-3m_b^4m_s^2-m_0^2m_b^3m_s+m_0^2m_bm_s^3}{6r^2r^{'2}}\\
&&+\frac{m_0^2m_s^2q^2-m_0^2m_b^2q^2+3m_b^2m_s^2q^2-3m_s^4q^2}{6r^2r^{'2}}\\
&&+\frac{m_0^2m_s^2+3m_sm_b^3+m_0^2q^2-3m_s^2m_b^2+3m_s^4-3m_s^2q^2}{6r^2r'}\\
&&+\frac{m_0^2m_bm_s-3m_s^3m_b}{6r^2r'}\\
&&+\frac{m_0^2m_b^4-2m_b^4m_s^2+2m_b^2m_s^4-m_0^2m_b^2m_s^2}{2r^3r'}\\
&&+\frac{2m_s^6-m_0^2m_s^4-2m_b^2m_s^4+m_0^2m_b^2m_s^2}{2rr^{'3}}\Bigg),
\end{eqnarray*}
where, $r=p^2-m_b^2$ and $r'=p^{'2}-m_s^2$.

\clearpage
\begin{figure}[th]
\vspace*{4.cm}
\begin{center}
\begin{picture}(160,20)
\put(-25,-50){ \epsfxsize=6.5cm
\epsfbox{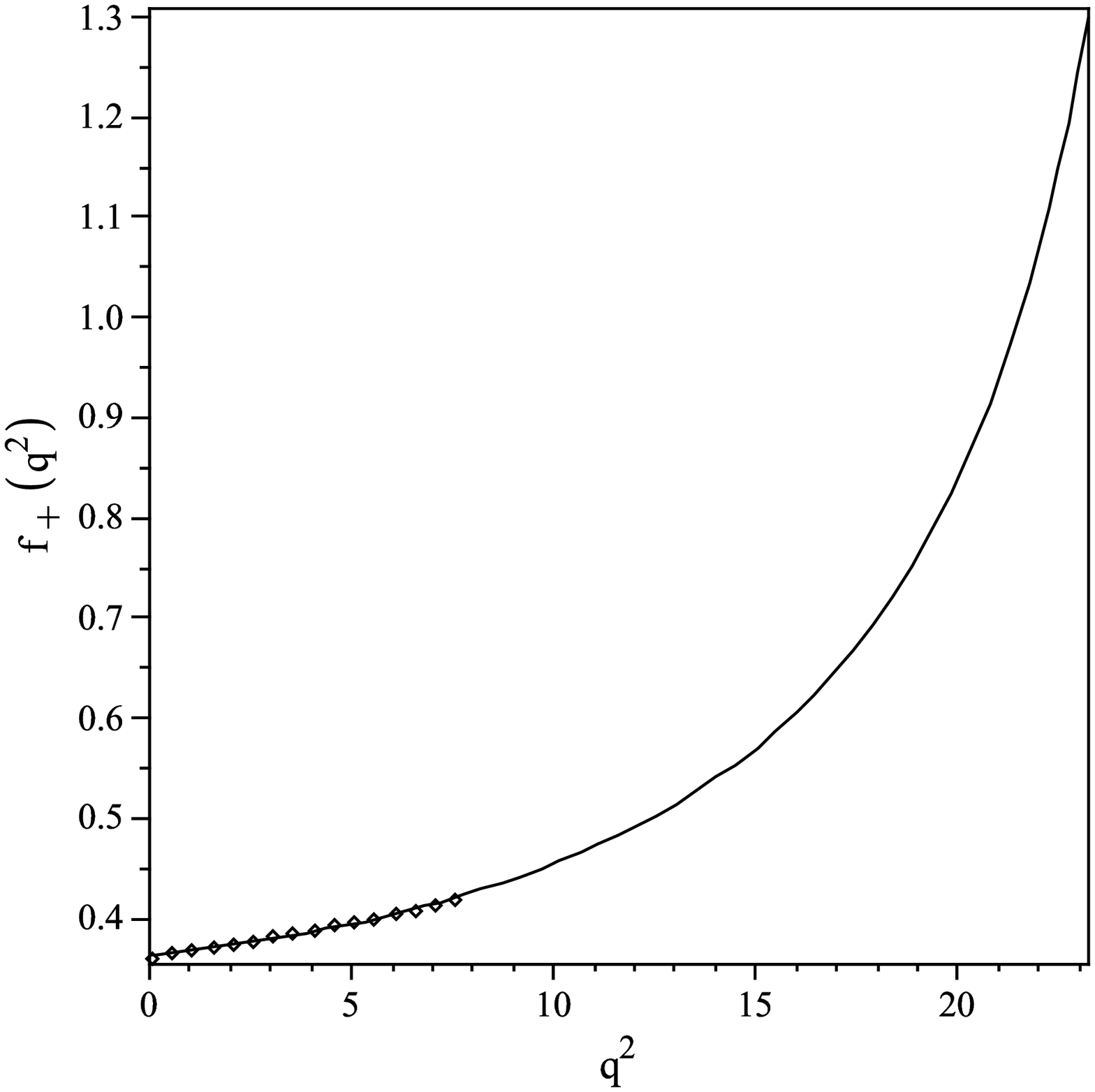}\epsfxsize=6.5cm \epsfbox{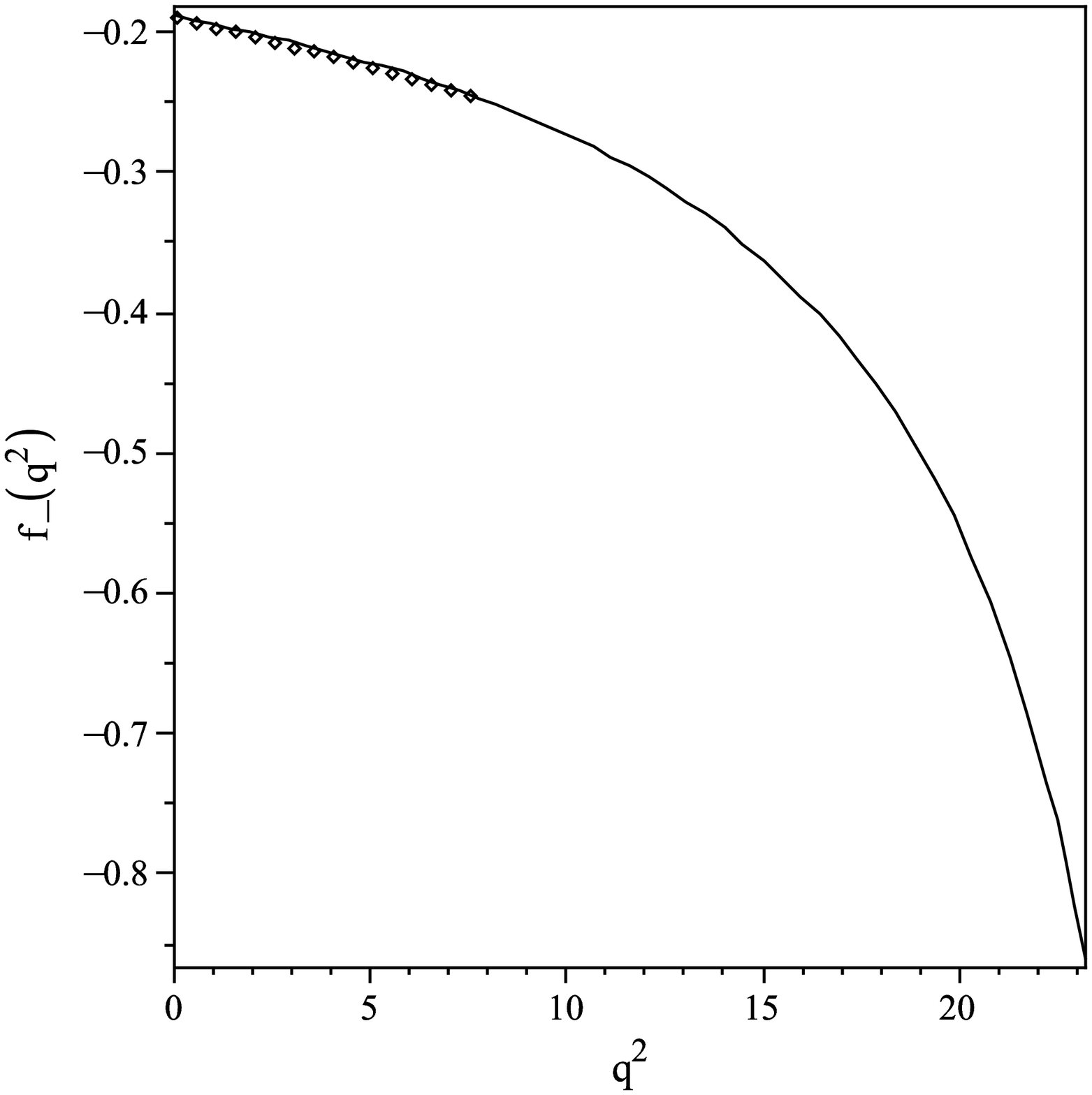}
\epsfxsize=6.5cm \epsfbox{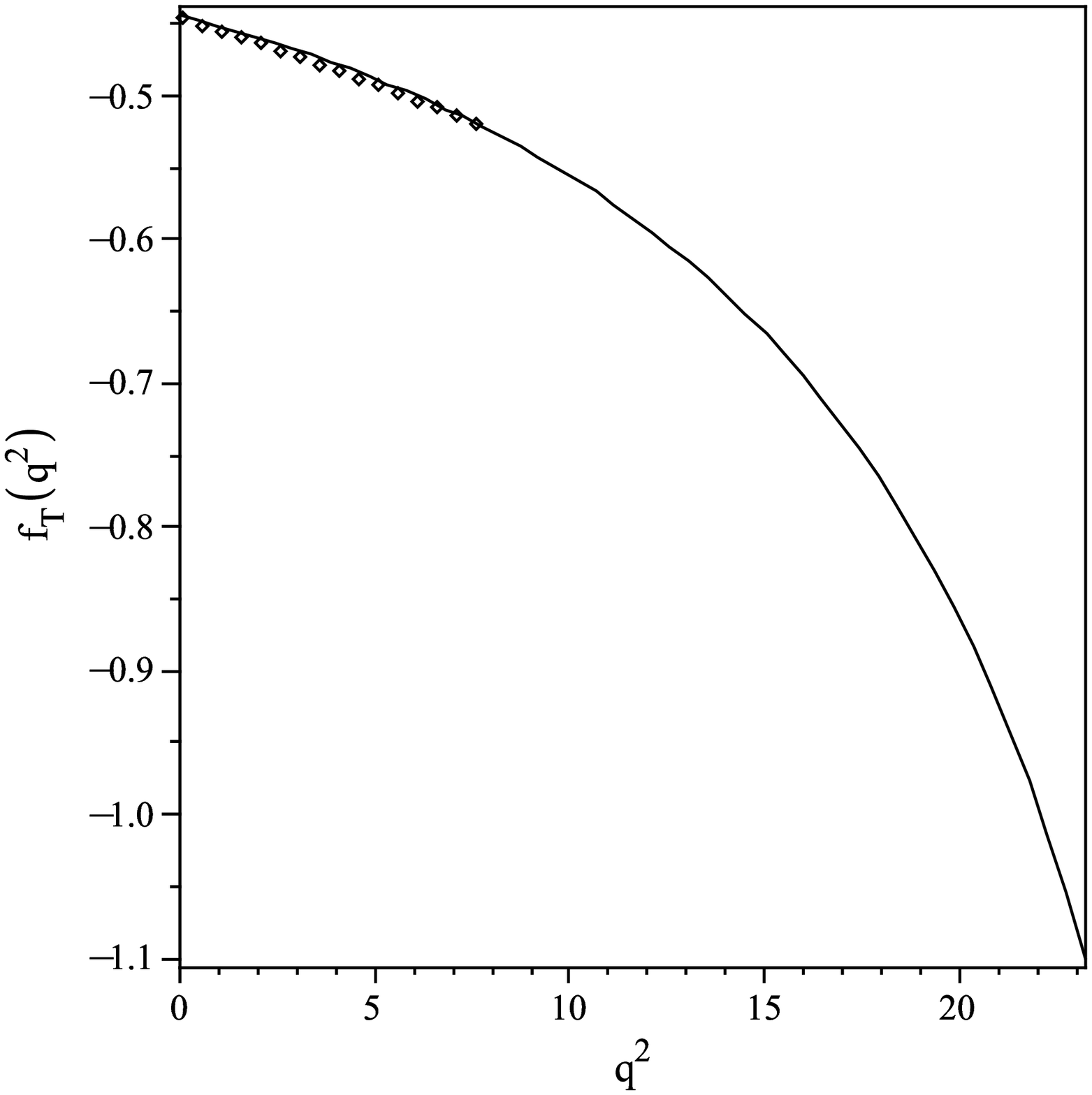}}
\end{picture}
\end{center}
\vspace*{5cm} \caption{The dependence of the form factors on $q^2$
at $M_1^2=12~GeV^2$ and $M_2^2=5GeV^2$ for  $P= \eta$. The small
boxes correspond to the values obtained directly from sum rules
and  the solid lines belong to the fit parametrization of the form
factors.}\label{F23}
\end{figure}
\newpage
\begin{figure}[th]
\vspace*{4.cm}
\begin{center}
\begin{picture}(160,20)
\put(-25,-50){ \epsfxsize=6.5cm
\epsfbox{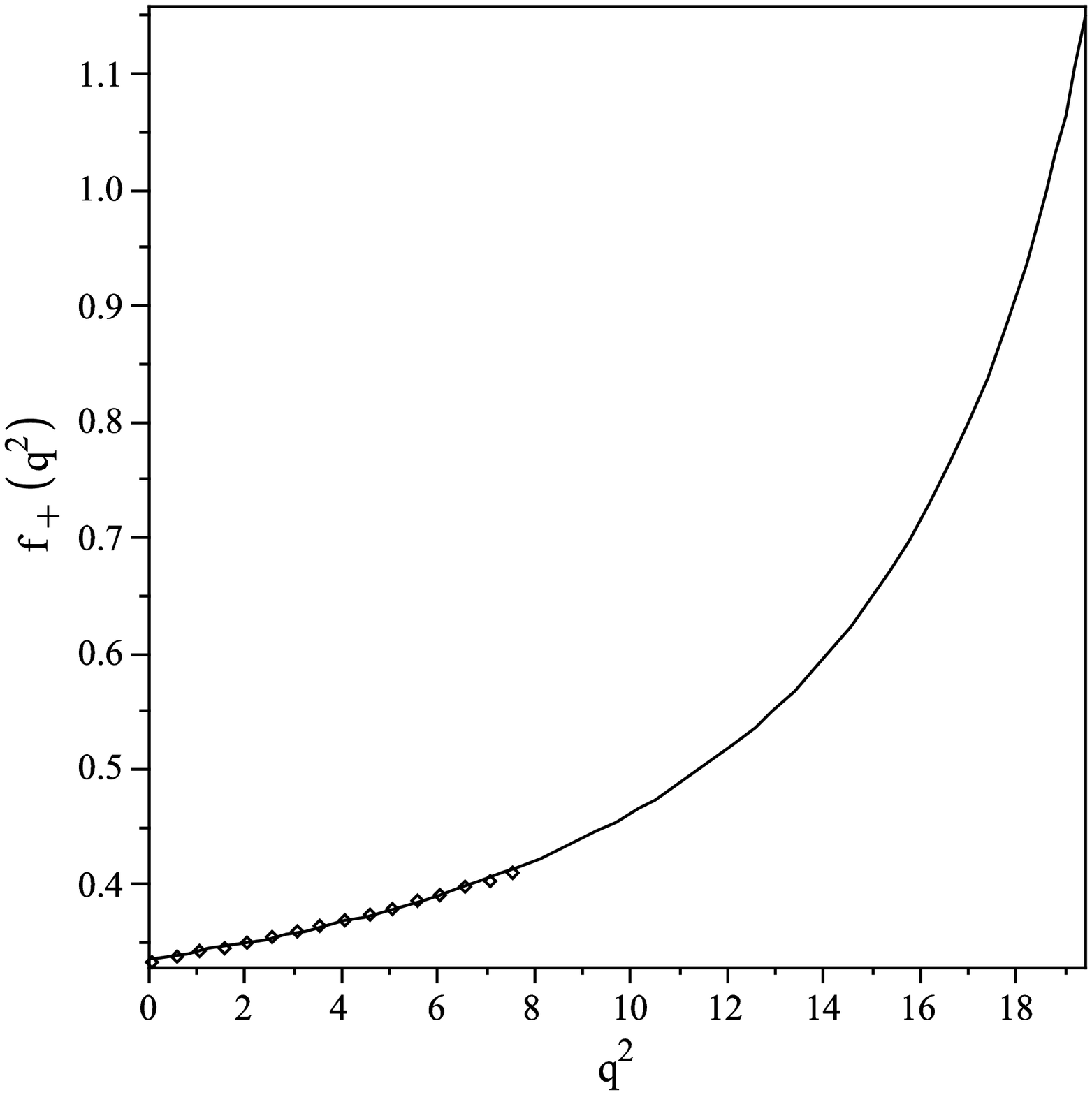}\epsfxsize=6.5cm \epsfbox{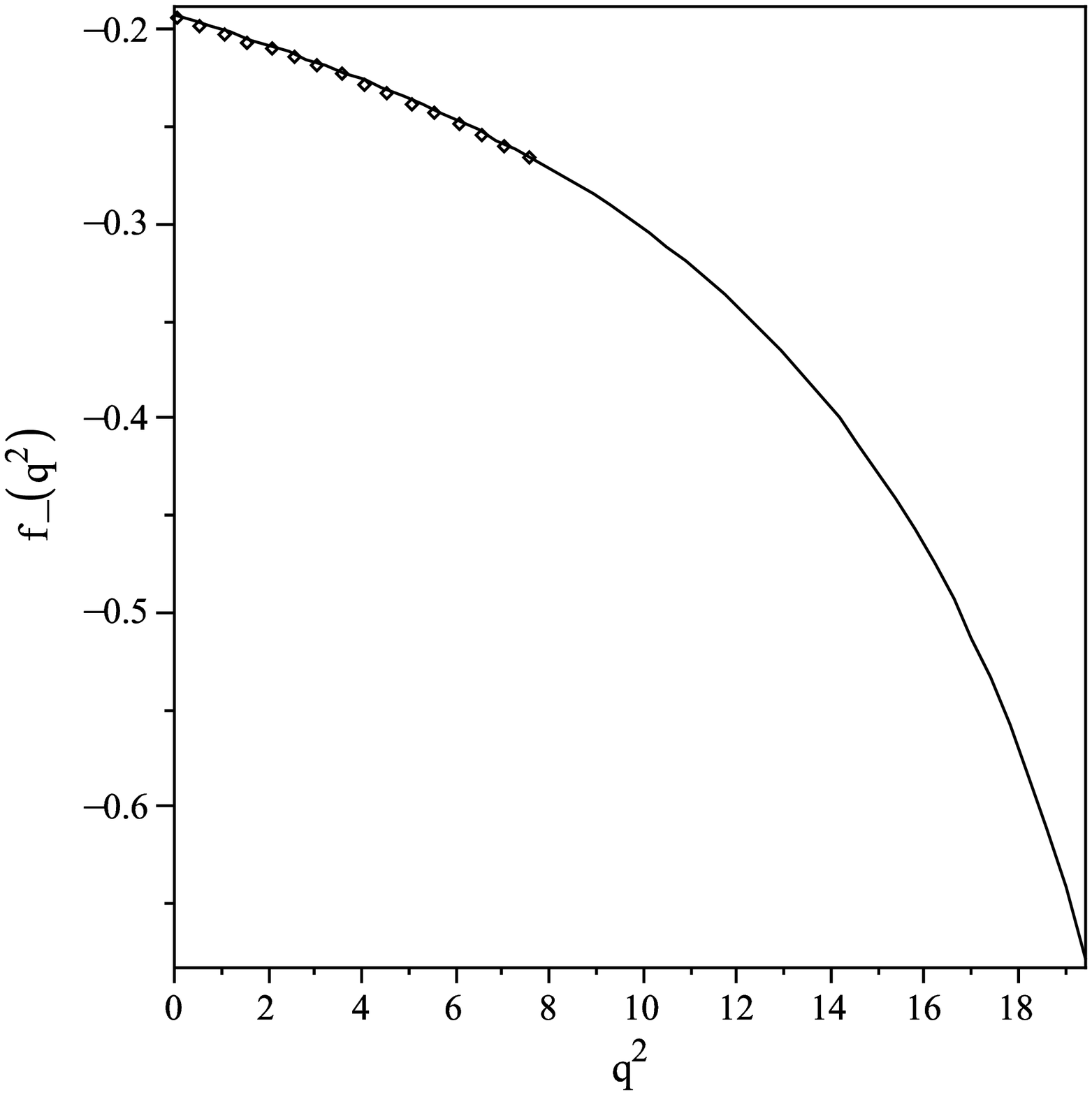}
\epsfxsize=6.5cm \epsfbox{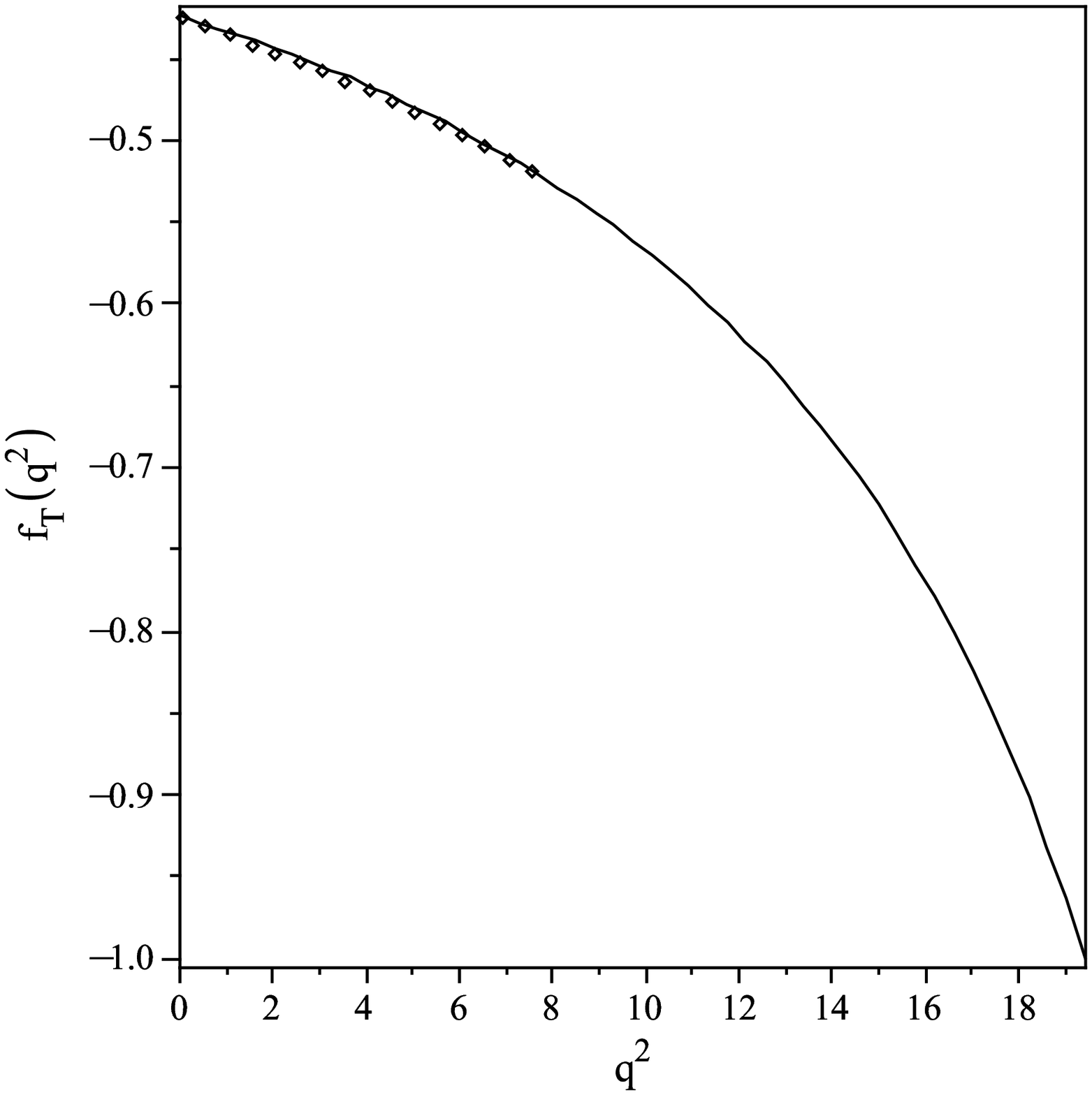}}
\end{picture}
\end{center}
\vspace*{5cm} \caption{The dependence of the form factors on $q^2$
at $M_1^2=12~GeV^2$ and  $M_2^2=5~GeV^2$ for  $P= \eta'$. The
small boxes correspond to the values obtained directly from sum
rules and  the solid lines  belong to the fit parametrization of
the form factors.}\label{F24}
\end{figure}
\newpage
\begin{figure}[th]
\begin{center}
\begin{picture}(160,100)
\centerline{ \epsfxsize=10cm \epsfbox{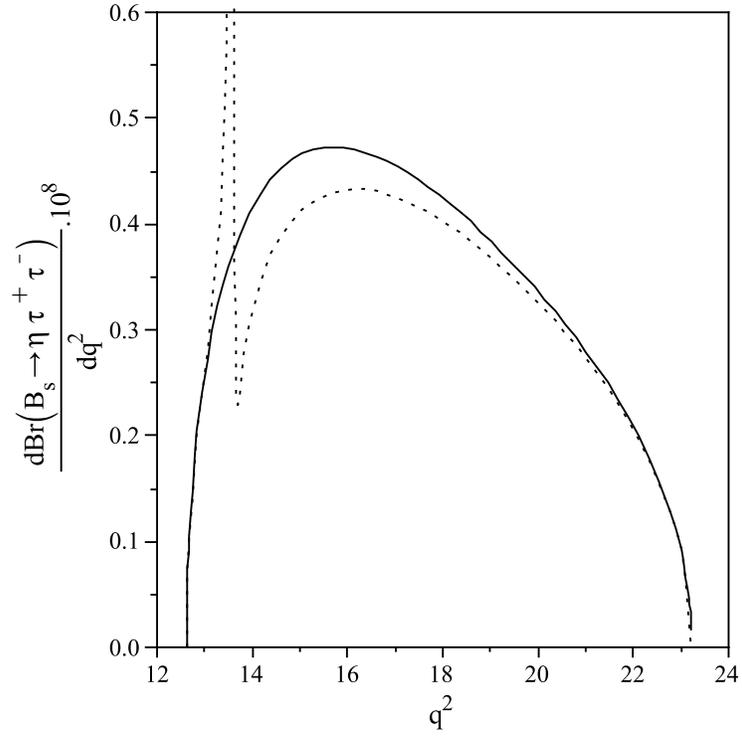}}
\end{picture}
\end{center}
\vspace*{0cm}\caption{The dependence of the differential
branching fraction of the $B_s\to \eta \tau^+\tau^-$ decay with
and without the LD effects  on $q^2$. The solid  and dotted lines
show the results without and with the LD effects,
respectively.}\label{F25}
\end{figure}
\normalsize
\newpage
\begin{figure}[th]
\begin{center}
\begin{picture}(160,100)
\centerline{ \epsfxsize=10cm \epsfbox{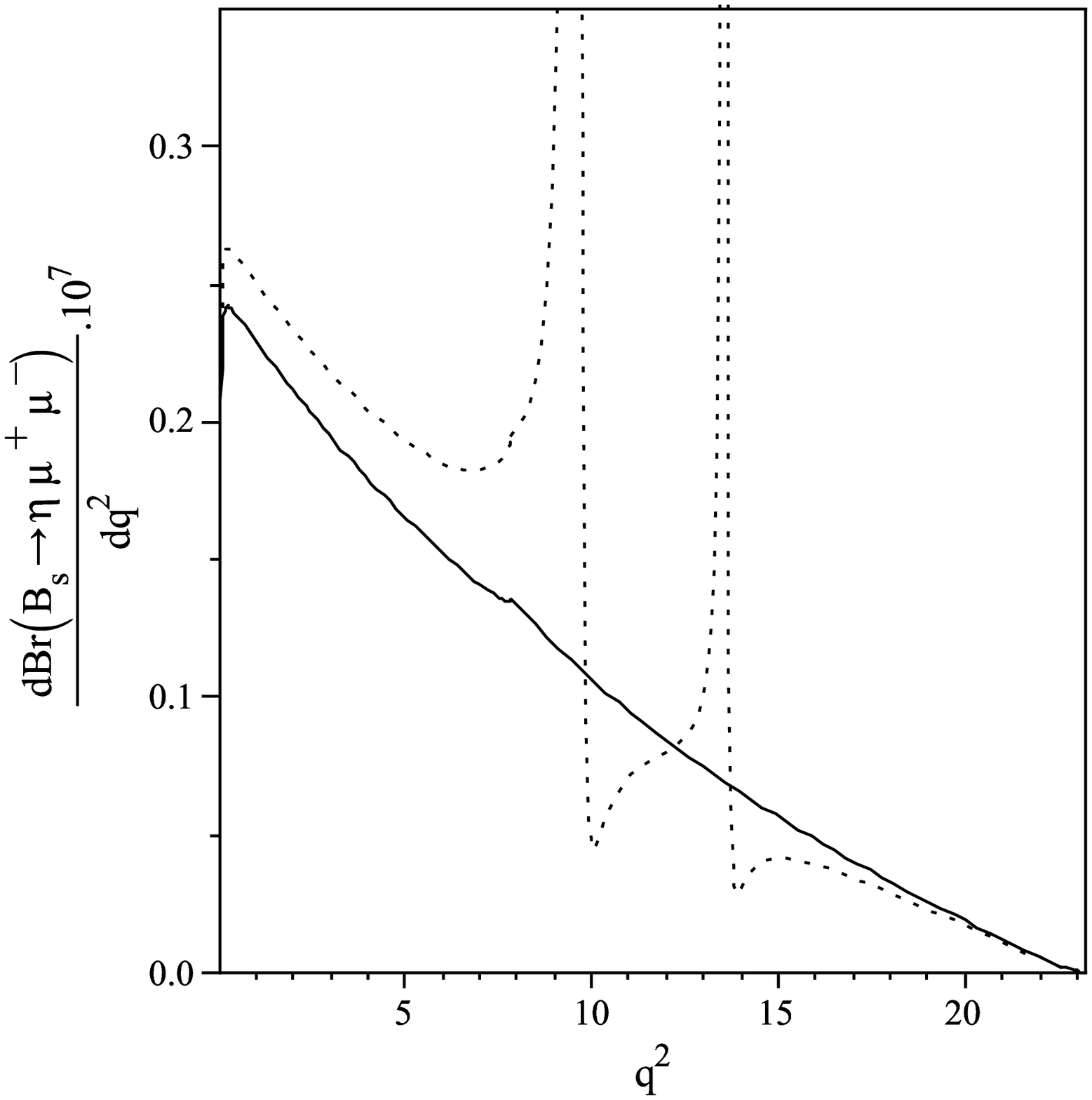}}
\end{picture}
\end{center}
\vspace*{0cm}\caption{The same as Fig \ref{F25} but for the
$B_s\to \eta \mu^+\mu^-$.}\label{F26}
\end{figure}
\normalsize
\newpage
\begin{figure}[th]
\begin{center}
\begin{picture}(160,100)
\centerline{ \epsfxsize=10cm \epsfbox{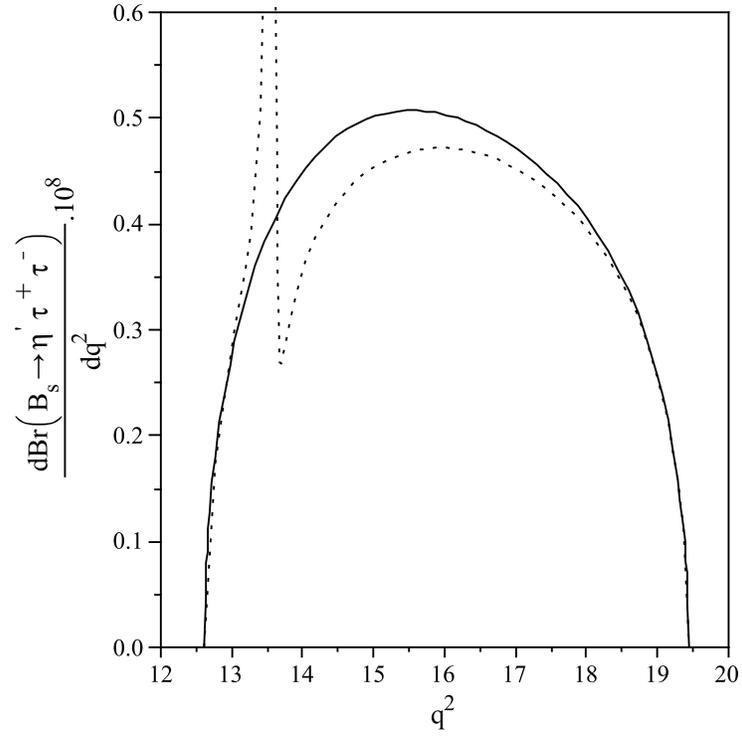}}
\end{picture}
\end{center}
\vspace*{0cm}\caption{The same as Fig \ref{F25} but for the
$B_s\to \eta' \tau^+\tau^-$.}\label{F28}
\end{figure}
\normalsize
\newpage
\begin{figure}[th]
\begin{center}
\begin{picture}(160,100)
\centerline{ \epsfxsize=10cm \epsfbox{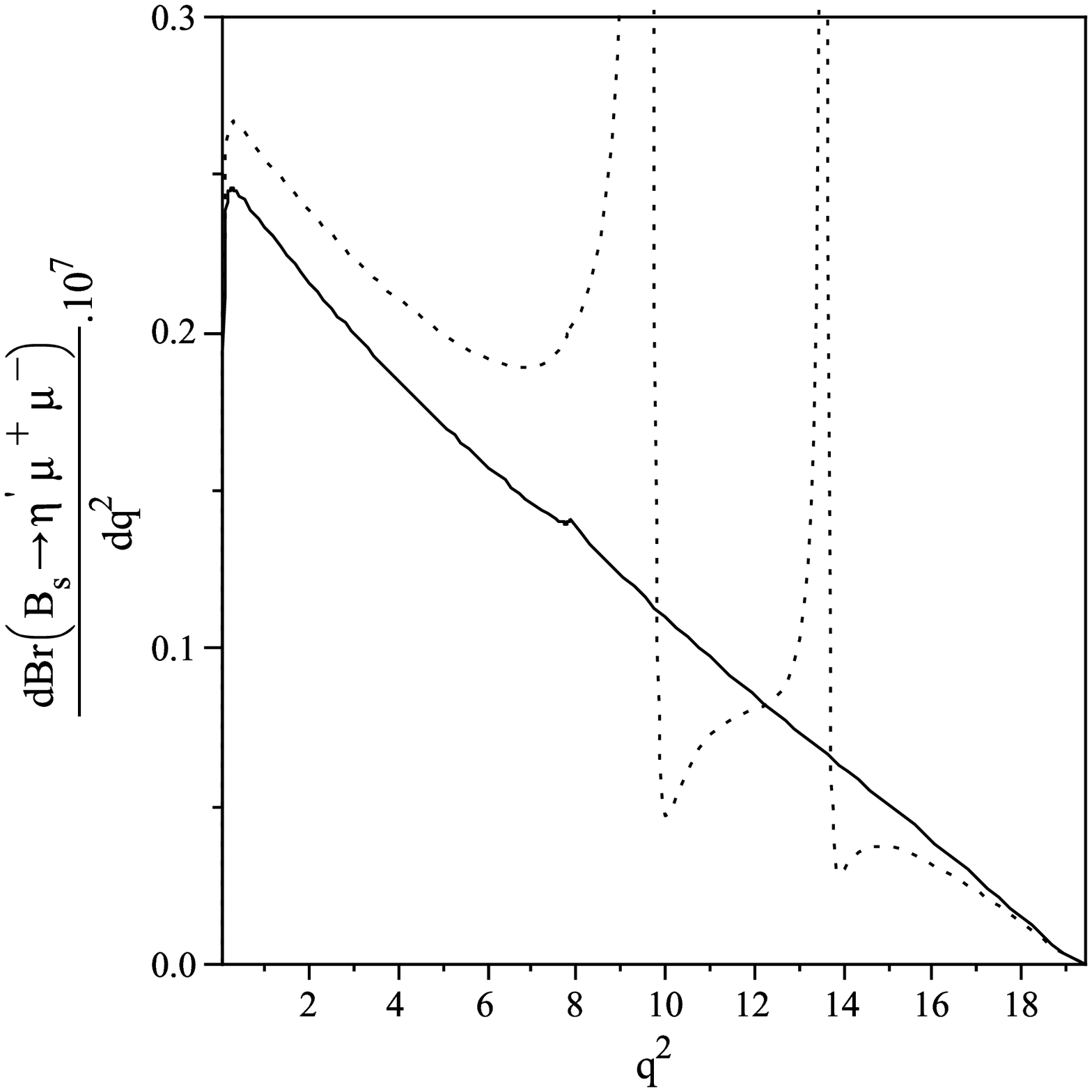}}
\end{picture}
\end{center}
\vspace*{0cm}\caption{The same as Fig \ref{F25} but for the
$B_s\to \eta' \mu^+\mu^-$.}\label{F29}
\end{figure}
\normalsize
\newpage
\begin{figure}[th]
\begin{center}
\begin{picture}(160,100)
\centerline{ \epsfxsize=10cm \epsfbox{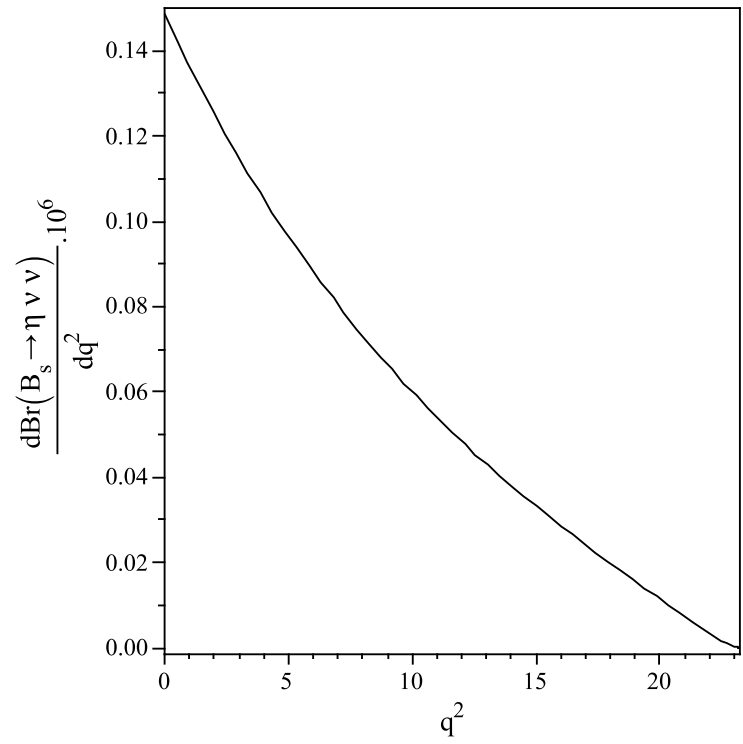}}
\end{picture}
\end{center}
\vspace*{0cm}\caption{The dependence of the differential
branching fraction of the $B_s\to \eta \nu\bar{\nu}$ decay on
$q^2$.}\label{F27}
\end{figure}
\normalsize
\newpage
\begin{figure}[th]
\begin{center}
\begin{picture}(160,100)
\centerline{ \epsfxsize=10cm \epsfbox{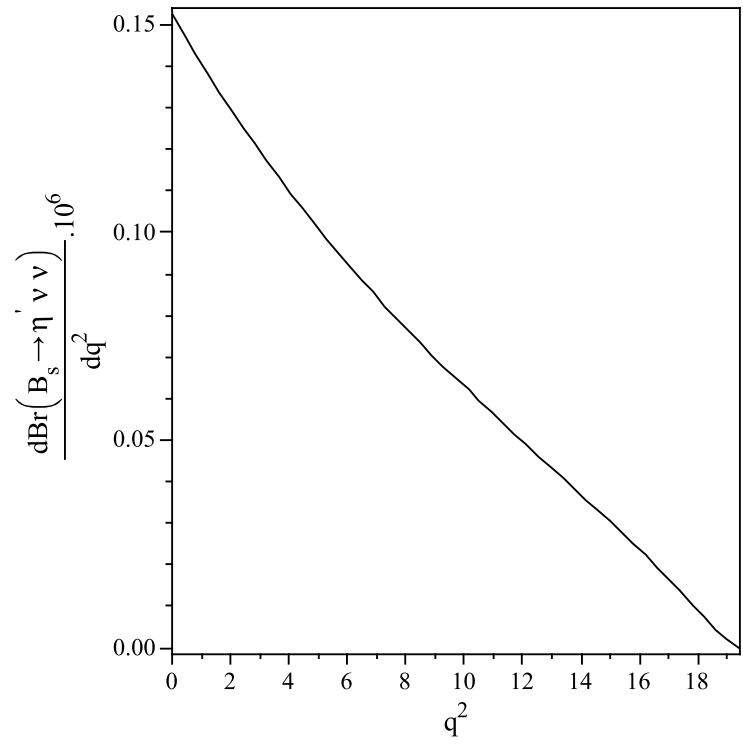}}
\end{picture}
\end{center}
\vspace*{0cm}\caption{The same as Fig \ref{F27} but for the
$B_s\to \eta' \nu\bar{\nu}$.}\label{F210}
\end{figure}
\normalsize
\newpage
\begin{figure}[th]
\begin{center}
\begin{picture}(160,20)
\put(-10,-110){ \epsfxsize=8cm \epsfbox{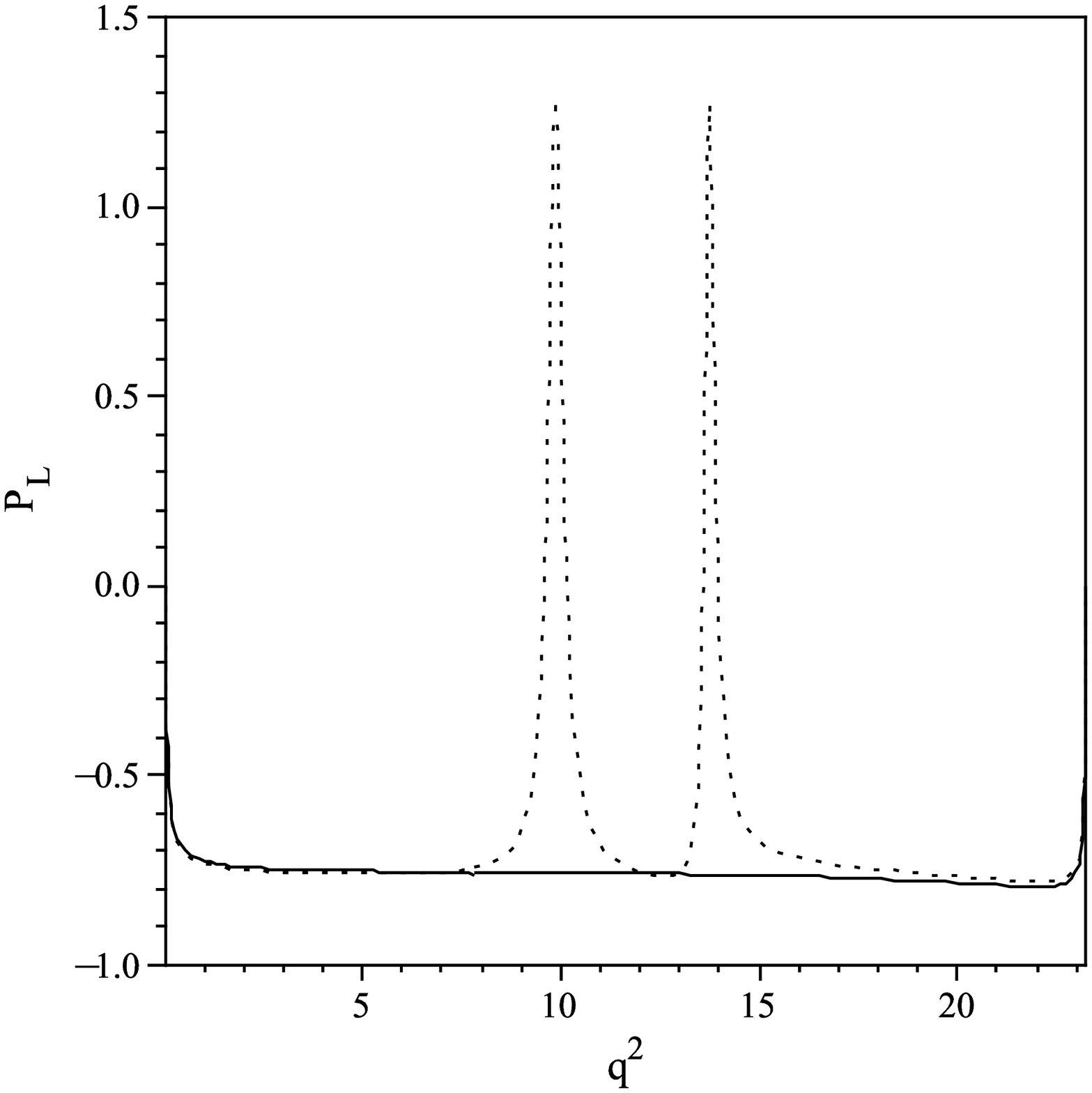} \epsfxsize=8cm
\epsfbox{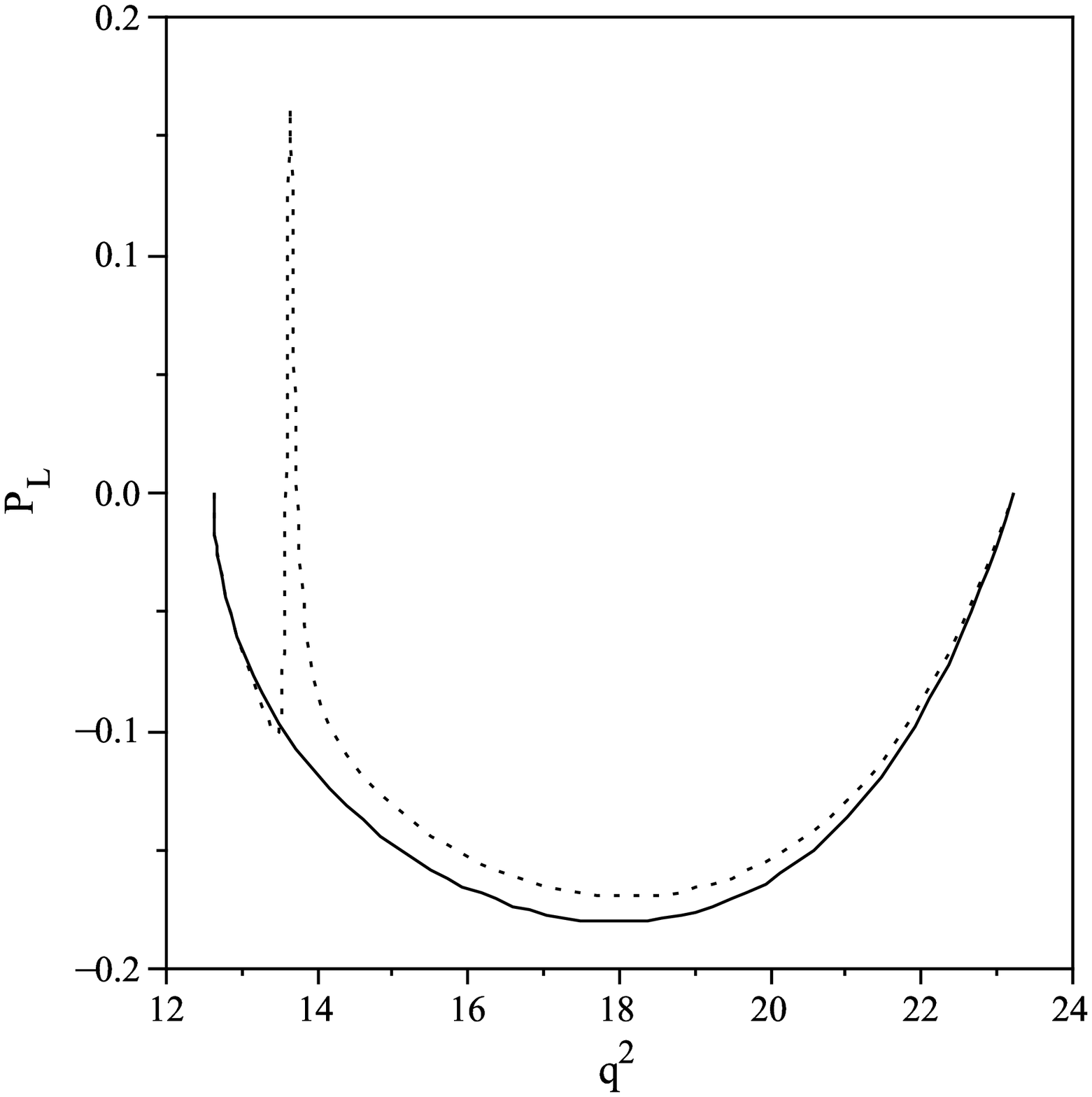}}
\end{picture}
\end{center}
\vspace*{11cm}\caption{The dependence of the Longitudinal lepton
polarization asymmetry on $q^2$. The left figure belongs to the
$B_s\to \eta \mu^+\mu^-$ decay and the right figure corresponds to
the $B_s\to \eta \tau^+\tau^-$. The solid lines and dotted lines
show the results without and with the  LD effects,
respectively.}\label{F211}
\end{figure}
\normalsize
\newpage
\begin{figure}[th]
\begin{center}
\begin{picture}(160,20)
\put(-10,-110){ \epsfxsize=8cm \epsfbox{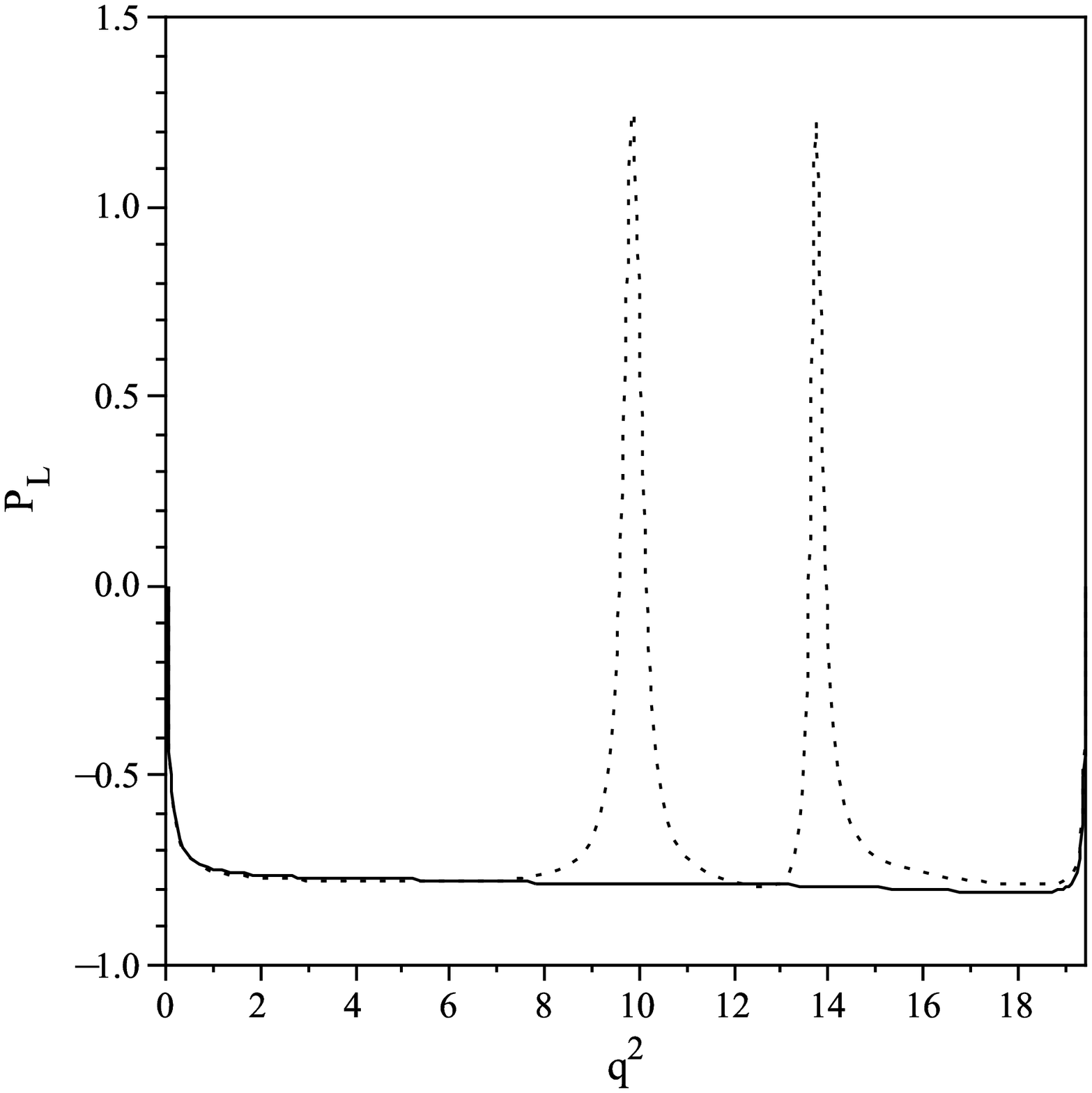} \epsfxsize=8cm
\epsfbox{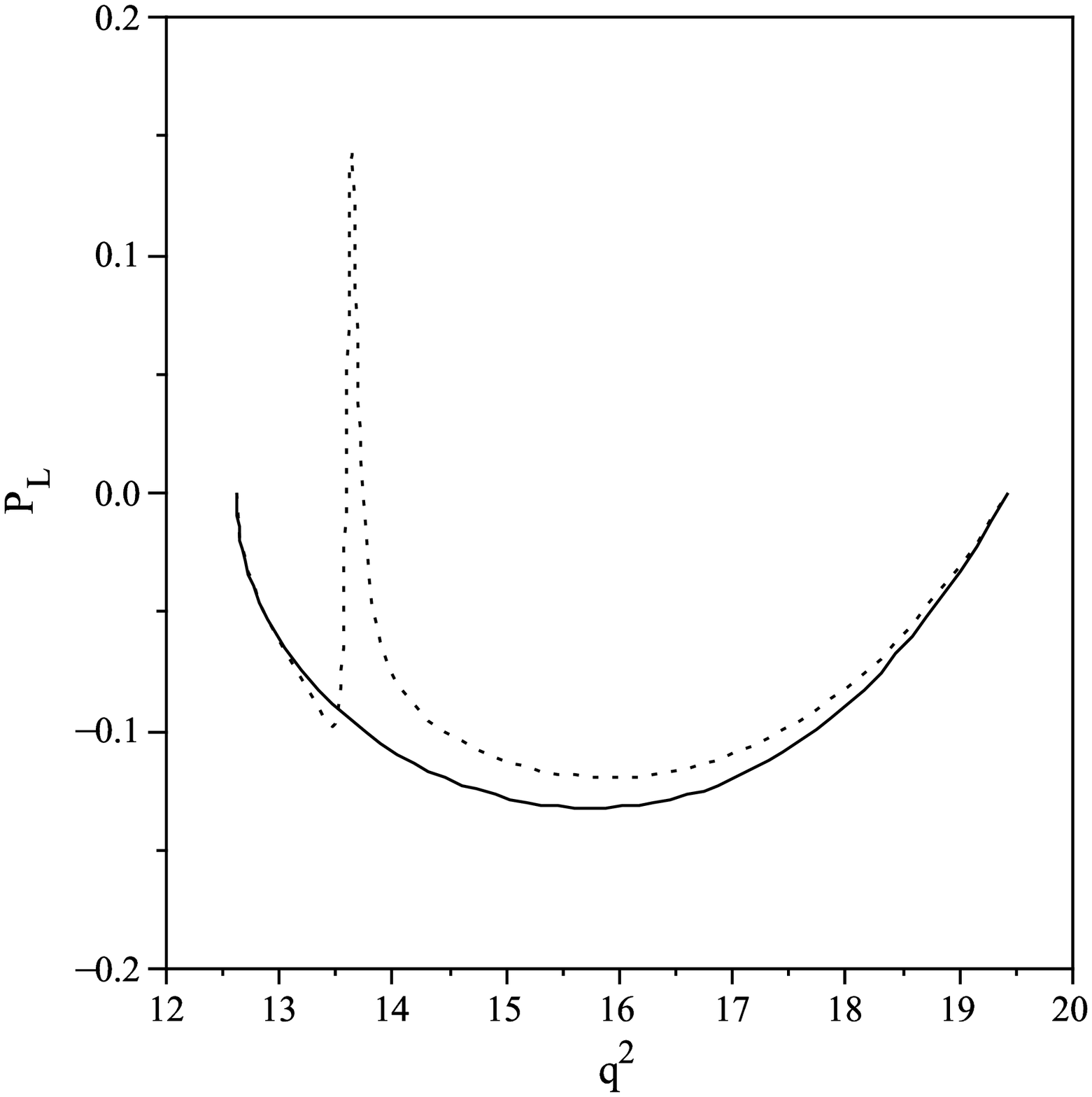}}
\end{picture}
\end{center}
\vspace*{11cm}\caption{The same as Fig \ref{F211} but for the $B_s
\to \eta'$ transition.}\label{F212}
\end{figure}


\normalsize
\newpage


\clearpage
\begin{thebibliography}{II}
\bibitem{CLEO05}
M. Artuso et al., (CLEO Collaboration), Phys. Rev. Lett. 95,
261801 (2005).

\bibitem{CLEO06}
G. Bonvicini et al., (CLEO Collaboration), Phys. Rev. Lett. 96,
022002 (2006).

\bibitem{Belle}
A. Drutskoy, arXiv:0905.2959 [hep-ex].

\bibitem{susy}
G. Buchalla, G. Hiller, G. Isidori, Phys. Rev. D 63, 014015
(2000).

\bibitem{dmat}
C. Bird, P. Jackson, R. Kowalewski, M. Pospelov, Phys. Rev. Lett.
93, 201803 (2004).

\bibitem{FKS}
T. Feldmann, P. Kroll,  B. Stech, Phys. Rev. D 58, 114006 (1998);
Phys. Lett. B 449, 339 (1999); T. Feldmann, Int. J. Mod. Phys. A
15, 159 (2000).

\bibitem{DP}
F. De Fazio, M. R. Pennington, JHEP 0007, 051 (2000).

\bibitem{HMC}
H. M. Choi, J. Phys. G 37, 085005 (2010).

\bibitem{CCD}
M. V. Carlucci, P. Colangelo, De. F. Fazio, Phys. Rev. D 80,
055023 (2009).

\bibitem{KLOE}
F. Ambrosino et al. (KLOE Collaboration), Phys. Lett. B 648, 267
(2007).

\bibitem{PDG}
C. Amsler et al., Particle Data Group, Phys. Lett. B 667, 1
(2008).

\bibitem{Buras}
A. J. Buras, M. Muenz, Phys. Rev. D 52, 186 (1995).

\bibitem{Rolf}
J. Rolf, M. Della Morte, S. Durr, J. Heitger, A. Juttner, H.
Molke, A. Shindler, R. Sommer, Nucl. Phys. Proc. Suppl. 129, 322
(2004).

\bibitem{Ball}
P. Ball, R. Zwicky,  Phys. Rev. D 71, 014015 (2005).

\bibitem{Chen}
C. H. Chen, C. Q. Geng, C. C. Lih, C. C. Liu, Phys. Rev. D 75,
074010 (2007).

\bibitem{GL}
C. Q. Geng, C. C. Liu, J. Phys. G 29, 1103 (2003).

\bibitem{Faessler}
A. Faessler, Th. Gutsche, M. A. Ivanov, J. G. K\"{o}rner, V. E.
Lyubovitskij,  Eur. Phys. J. C 4, 18 (2002).

\end{thebibliography}
\end {document}